\documentclass[twocolumn,10pt,letterpaper]{article}
\usepackage[top=2cm, bottom=2cm, left=1.8cm, right=1.8cm]{geometry}
\usepackage{graphicx}
\usepackage{float}
\usepackage{paralist}
\usepackage{times}
\usepackage[hyphens]{url}

\usepackage{booktabs} %
\usepackage{graphicx}
\usepackage{float}

\usepackage[bf]{caption}
\usepackage{subcaption}
\usepackage{afterpage}
\usepackage{float}
\usepackage{amsmath}
\usepackage{amsfonts}
\usepackage{color, colortbl}

\usepackage{cleveref}

\usepackage{paralist}

\renewenvironment{enumerate}[1]{\begin{compactenum}#1}{\end{compactenum}}

\usepackage[marginal,hang,stable]{footmisc}

\renewcommand{\footnoterule}{%
  \kern -3pt
  \hrule width 1in 
  \kern 2pt
}

\makeatletter
\def\url@leostyle{%
  \@ifundefined{selectfont}{\def\UrlFont{}}%
  {\def\UrlFont{}}%
}
\makeatother
  \newcommand\suba{0.323}
  \newcommand\subb{0.42}
  \newcommand\subc{0.4}
  \newcommand\sube{0.32}
  \newcommand{\descr}[1]{\smallskip \noindent \textbf{#1}}

\newcommand{\WB}{\ensuremath{\mbox{$\mathcal{A}_{wb}$}}~}
\newcommand{\BB}{\ensuremath{\mbox{$\mathcal{A}_{bb}$}}~}

\let\OLDthebibliography\thebibliography
\renewcommand\thebibliography[1]{
  \OLDthebibliography{#1}
  \setlength{\parskip}{0.5pt}
  \setlength{\itemsep}{0pt plus 0.3ex}
}

\usepackage{titlesec}
\titlespacing*{\section}{0pt}{*2}{5pt} 
\titlespacing{\subsection}{0pt}{*2}{4pt}
\titlespacing{\subsubsection}{0pt}{*1.5}{3pt}

\usepackage{etoolbox} 
\makeatletter
\patchcmd\maketitle{\@makefntext}{\@@@ddt}{}{}
\patchcmd\maketitle{\rlap}{\mbox}{}{}
\makeatother

\definecolor{darkgreen}{RGB}{135,0,40}
\usepackage[bookmarks=false, colorlinks=true, plainpages = false,  linkcolor = darkgreen, citecolor = darkgreen, urlcolor = darkgreen, filecolor = black]{hyperref}

\begin{document}
\title{\bf LOGAN: Membership Inference Attacks Against Generative Models\thanks{Published in the Proceedings on Privacy Enhancing Technologies (PoPETs), Vol. 2019, Issue 1.}}
\author{Jamie Hayes\thanks{Authors contributed equally.}, Luca Melis$^\dag$, George Danezis, and Emiliano De Cristofaro\\[1ex]
\normalsize University College London\\
\normalsize {\{j.hayes, l.melis, g.danezis, e.decristofaro\}@cs.ucl.ac.uk}}
\date{}

\maketitle

\begin{abstract}
Generative models estimate the underlying distribution of a dataset to generate realistic samples according to that distribution. In this paper, we present the first membership inference attacks against generative models: given a data point, the adversary determines whether or not it was used to train the model. Our attacks leverage Generative Adversarial Networks (GANs), which combine a discriminative and a generative model, to detect overfitting and recognize inputs that were part of training datasets, using the discriminator's capacity to learn statistical differences in distributions.

We present attacks based on both white-box and black-box access to the target model, against several state-of-the-art generative models, over datasets of complex representations of faces (LFW), objects (CIFAR-10), and medical images (Diabetic Retinopathy). We also discuss the sensitivity of the attacks to different training parameters, and their robustness against mitigation strategies, finding that defenses are either ineffective or lead to significantly worse performances of the generative models in terms of training stability and/or sample quality. 

\end{abstract}

\section{Introduction}\label{sec:intro}
Over the past few years, providers such as  Google, Microsoft, and Amazon have started to provide customers with access to APIs allowing them to easily embed machine learning tasks into their applications.
Organizations can use Machine Learning as a Service (MLaaS) engines to outsource complex tasks, e.g., training classifiers, performing predictions, clustering, etc. %
They can also let others query models trained on their data, possibly at a cost.
However, if malicious users were able to recover data used to train these models, the resulting information leakage would create serious issues.
In particular, organizations do not have much control over the kind of models and training parameters used by the %
platform,  and this might lead to overfitting (i.e., the model does not generalize well outside the data on which it was trained), which provides attackers with a useful tool to recover training data~\cite{shokri2016membership}.

In recent years, research in deep learning has made tremendous progress in the area of {\bf generative models}.
These models are used to generate new samples from the same underlying distribution of a given training dataset.
In particular, generative models offer a way to artificially generate plausible images and videos and they are used in many applications, e.g., compression~\cite{theis2017lossy}, denoising~\cite{bengio2013generalized}, inpainting~\cite{yeh2016semantic}, super-resolution~\cite{ledig2016photo}, semi-supervised learning~\cite{salimans2016}, etc. 

In this paper, we study the feasibility of {\bf membership inference attacks} against generative models.
That is, given access to a generative model and an individual data record, can an attacker tell if a specific record was used to train the model?
Membership inference on generative models is likely to be more challenging than on discriminative ones (see, e.g.,~\cite{shokri2016membership}). %
The latter attempt to predict a label given a data input, and an attacker can use the confidence the model places on an input belonging to a label to perform the attack. In generative models, there is no such signal, thus, it is difficult to  both detect overfitting and infer membership.

\subsection{Motivation}

We study how generating synthetic samples through generative models may lead to information leakage.
In particular, we focus on membership inference attacks against them, which are relevant to, and can be used in, a number of settings: %

\descr{Direct privacy breach.} Membership inference can directly violate privacy if inclusion in a training set is itself sensitive. 
For example, if synthetic health-related images (i.e., generated by generative models) are used for research purposes, discovering that a specific record was used for training leaks information about the individual's health.  (Note that image synthesis is commonly used to create datasets for healthcare applications~\cite{nie2016medical, choi17generating}.) %
Similarly, if images from a database of criminals are used to train a face generation algorithm~\cite{wu2016automated}, membership inference may expose an individual's criminal history. 

\descr{Establishing wrongdoing.} Regulators can use membership inference to support the suspicion that a model was trained on personal data without an adequate legal basis, or for a purpose not compatible with the data collection.
For instance, DeepMind was recently found to have used personal medical records provided by the UK's National Health Service for purposes beyond direct patient care; the basis on which the data was collected~\cite{verge}.
In general, membership inference against generative models allow regulators to assess whether personal information has been used to train a generative model. 

\descr{Assessing privacy protection.} Our methods can be used by cloud providers that offer MLaaS for generative models (e.g., Neuromation\footnote{\url{https://neuromation.io}}) to evaluate the level of ``privacy'' of a trained model. 
In other words, they can use them as a benchmark before allowing third parties access to the model; providers may restrict access in case the inference attack yields good results. 
Also, susceptibility to membership inference likely correlates with other leakage and with overfitting; in fact, the relationship between robust privacy protections and generalizations have been discussed by Dwork et al.~\cite{dwork2015generalization}.

\smallskip\noindent Overall, membership inference attacks are often a gateway to further attacks. That is, the adversary first infers whether data of a victim is part of the information she has access to (a trained model in our case), and then mount other attacks (e.g., profiling~\cite{pyrgelis2017does}, property inference~\cite{ateniese2015hacking,melis2018inference}, etc.), which might leak additional information about the victim.

\subsection{Roadmap}

\noindent\textbf{Attacks Overview.} We consider both black-box and white-box attacks: in the former, the adversary can only make queries to the model under attack, i.e., the \emph{target model}, and has no access to the internal parameters. In the latter, he also has access to the parameters. %
To mount the attacks, %
we train a Generative Adversarial Network (GAN) model~\cite{goodfellow2014generative} on samples generated from the target model;
specifically, we use generative models as a method to learn information about the target generative model, and thus create a local copy of the target model from which we 
can launch the attack.
Our intuition is that, if a generative model overfits, then a GAN, which combines a discriminative model and a generative model, should be able to detect this overfitting, even if it is not observable to a human, since the discriminator is trained to learn statistical differences in distributions. We rely on GANs to classify real and synthetic records to recognize differences in samples generated from the target model, on inputs on which it was trained versus those on which it was not.
Moreover, for white-box attacks, the attacker-trained discriminator itself can be used to measure information leakage of the target model.

\descr{Experiments.} We test our attacks on several state-of-the-art models: Deep Convolutional GAN (DCGAN)~\cite{radford2015unsupervised}, Boundary Equilibrium GAN (BEGAN)~\cite{berthelot2017began}, and the combination of DCGAN with a Variational Autoencoder (DCGAN+VAE)~\cite{larsen2015autoencoding}, using datasets with complex representations of faces (LFW), objects (CIFAR-10), and medical images (Diabetic Retinopathy), containing rich details both in the foreground and background. 
This represents a much more challenging task for the attacker compared to simple datasets such as MNIST, where samples from each class have very similar features.

\descr{Contributions.} In summary, our contributions include:\smallskip
\begin{compactenum}
\item We present the first study of membership inference attacks on generative models;\smallskip
\item We devise a white-box attack that is an excellent indicator of overfitting in generative models,
and a black-box attack that can be mounted through Generative Adversarial Networks, and show how to boost the performance of the black-box attack via auxiliary attacker knowledge of training/testing set;\smallskip
\item We show that our white-box attacks are 100\% successful at inferring which samples were used to train the target model, while we can recover up to over 80\% of the training set with black-box access; \smallskip
\item We investigate possible defense strategies, including training regularizers, %
showing that they are either ineffective or lead to significantly worse performances of the models in terms of the quality of the samples generated and/or training stability.
\end{compactenum}

\descr{Paper Organization.} The rest of this paper is organized as follows. 
The next section reviews related work, then, Section~\ref{sec:background} introduces machine learning concepts used in the rest of the paper, while Section~\ref{sec:outline} presents our attacks.
In Section~\ref{sec:experiments}, we present the results of our experimental evaluation, and, in Section~\ref{sec:discussion}, we discuss the cost of our attacks as well as  possible mitigation strategies.
Finally, the paper concludes in Section~\ref{sec:conclusion}.

\section{Related Work}\label{sec:related}
We now review prior work on attacks and defense mechanisms on machine learning models.

\subsection{Attacks}

Over the past few years, a few privacy attacks on machine learning have been proposed.
For instance, attacks targeting distributed recommender systems~\cite{calandrino2011you} have focused on inferring which inputs cause output changes by looking at temporal patterns of the model.

Specific to membership inference are attacks against supervised models by Shokri et al.~\cite{shokri2016membership}. Their approach exploits 
differences in the model's response to inputs that were or were not seen during training. For each class of the targeted black-box model, they train a \emph{shadow model}, with the same machine learning technique.
Whereas, our approach targets generative models and relies on GANs to provide a general framework for measuring the information leakage. 
As mentioned earlier, membership inference on generative models is much more challenging than on discriminative models: in the former, the attacker cannot exploit confidence values on inputs belonging to the same classes, thus it is more difficult to detect overfitting and mount the attack.
As a matter of fact, detecting overfitting in generative models is regarded as one of the most important research problems in machine learning~\cite{wu2016quantitative}.
Overall, our work presents black-box attacks that do not rely on any prediction vectors from the target model, as generic generative models output synthetic samples.

Additional membership inference attacks focus on genomic research studies~\cite{homer2008resolving, backes2016membership}, whereby an attacker aims to infer the presence of a particular individual's data within an aggregate genomic dataset, or aggregate locations~\cite{pyrgelis2017knock}.

Then, in {\em model inversion} attacks~\cite{fredrikson2014privacy}, an adversary extracts training data from outputted model predictions. %
Fredrikson et al.~\cite{fredrikson2015model} show how an attacker can rely on outputs from a model to infer sensitive features used as inputs to the model itself: given the model and some demographic information about a patient whose records are used for training, an attacker predicts  sensitive attributes of the patient.
However, the attack %
does not generalize on inputs not seen at training time, thus, the attacker relies on statistical inference about the total population~\cite{blogInversion}.
The record extracted by the attacker is not an actual training record, but an average representation of the inputs that are classified in a particular class. 
Long et al.~\cite{long2018understanding} and Yeom et al.~\cite{yeom2017unintended} investigate connections between membership inference and model inversion attacks against machine learning classifiers. 
In particular,~\cite{yeom2017unintended} assumes that the adversary knows the distribution from which the training set was drawn and its size, and that the adversary colludes with the training algorithm.
Their attacks are close in performance to Shokri et al.'s~\cite{shokri2016membership}, and show that, besides overfitting, the influence of target attributes on model's outputs also correlates with successful attacks.
Then, Tramer et al.~\cite{tramer2016stealing} present a model extraction attack to infer the parameters from a trained classifier, however, it only applies to scenarios where the attacker has access to the probabilities returned for each class.

Song et al.~\cite{song2017machine} attacks force a machine learning model to memorize the training data in such a way that an adversary can later extract training inputs with only black-box access to the model.
Then, Carlini et al.~\cite{carlini2018secret} show that deep learning-based language models trained on text data can unintentionally memorize specific training inputs, which can then be extracted with black-box access, however,  demonstrating it only for simple sequences of digits artificially introduced into the text.
Ateniese et al.~\cite{ateniese2015hacking} present a few attacks against SVM and HMM classifiers aimed to reconstruct properties about training sets, by exploiting knowledge of model parameters. 

Also, recent work~\cite{melis2018inference,aono2017privacy,hitaj2017deep} present inference attacks against distributed deep learning~\cite{shokri2015privacy,mcmahan2016communication}.
In particular, Aono et al.~\cite{aono2017privacy} target the collaborative privacy-preserving deep learning protocol of~\cite{shokri2015privacy}, and show that an honest-but-curious server can partially recover participants' data points from the shared gradient updates.
However, they operate on a simplified setting where the batch consists of a single data point.
Also, Hitaj et al.~\cite{hitaj2017deep} introduce a white-box attack against~\cite{shokri2015privacy}, which relies on GAN models to generate valid samples of a particular class from a targeted private training set, however, it cannot be extended to black-box scenarios.
Furthermore, evaluation of the attack is limited to the MNIST dataset of handwritten digits where all samples in a class look very similar, and the AT\&T Dataset of Faces,
which consists of only 400 grayscale images of faces. By contrast, our evaluation is performed on 13,233, 60,000, and 88,702 images for the LFW, CIFAR-10, and Diabetic Retinopathy datasets, respectively (see Section~\ref{sec:experiments}).

Finally, Truex et al.~\cite{demyst2018} show how membership inference attacks are data-driven and largely transferable, while Melis et al.~\cite{melis2018inference} demonstrate how an adversarial participant can successfully perform membership inference in distributed learning~\cite{shokri2015privacy,mcmahan2016communication}, as well as inferring sensitive properties that hold only for a subset of the participants' training data.

\subsection{Defenses}

Privacy-enhancing tools based on secure multiparty computation and homomorphic encryption have been proposed to securely train supervised machine learning models, such as decision trees~\cite{lindell2000privacy}, linear regressors~\cite{du2004privacy}, and neural networks~\cite{bonawitzpractical,dowlin2016cryptonets}.
However, these mechanisms do not prevent an attacker from running inference attacks on the privately trained models as the final parameters are left unchanged.    

Differential Privacy~\cite{dwork2008differential} can be used to mitigate inference attacks, and it has been widely applied to various machine learning models~\cite{abadi2016deep,papernot2016semi,shokri2015privacy,wainwright2012privacy,kusner2015differentially,wu2015revisiting}. 
Shokri and Shmatikov~\cite{shokri2015privacy} support distributed training of deep learning networks in a privacy-preserving way, where independent entities collaboratively build a model without sharing their training data,
but selectively share subsets of noisy model parameters during training. 
Abadi et al.~\cite{abadi2016deep} show how to train deep neural networks (DNNs) with non-convex objectives with an acceptable privacy budget,
while Rahman et al.~\cite{rahmanmembership} show that Abadi et al.'s proposal partially mitigates the effects of Shokri et al.'s~\cite{shokri2016membership} membership inference attack.

Papernot et al.~\cite{papernot2016semi,papernot2018scalable} combine %
multiple models trained with disjoint datasets without exposing the 
models, %
while, in~\cite{papernot2016distillation}, present ``defensive distillation'' to reduce the effectiveness of adversarial samples on DNNs.

Then, Beaulieu et al.~\cite{beaulieu2017privacy} apply the noisy gradient descent from~\cite{abadi2016deep} to train the discriminator of a Generative Adversarial Network under differential privacy. The resulting model is then used to generate synthetic subjects based on the population of clinical trial data.
Finally, Jia et al.\cite{jia2018attriguard} use adversarial machine learning to defend against attribute inference attacks, in the setting where an attacker trains a classifier to infer a target user's sensitive attributes from their public data, while Nasr et al.~\cite{nasr2018machine} leverage adversarial regularizers to design a privacy-preserving training mechanism with provable protections against membership inference attacks against discriminative models.

\section{Background}\label{sec:background}
In this section, we review machine learning concepts used throughout the paper.

\descr{Generative Models.} Machine learning models include discriminative and generative ones.
Given a supervised learning task, and given the features $(\mathbf{x})$ of a data-point and the corresponding label $(y)$, %
discriminative models attempt to predict $y$ on future $\bf{x}$ by learning a discriminative function $f$ from $(\bf{x},y)$; the function takes in input $\bf{x}$ and outputs the most likely label $y$.
Discriminative models are not able to ``explain'' how the data-points might have been generated.
By contrast, generative models describe how data is generated by learning the joint probability distribution of  $p(\mathbf{X}, \mathbf{Y})$, which gives a score to the configuration determined together by pairs $(\bf{x},y)$.
Generative models based on deep neural networks, such as Generative Adversarial Networks (GAN)~\cite{goodfellow2014generative} (introduced below) and Variational Auto-encoders (VAE)~\cite{kingma2013auto}  are considered the state-of-the-art for producing samples of realistic images~\cite{generative}. 

\begin{figure}[t]
    \centering
        \includegraphics[width=0.48\textwidth]{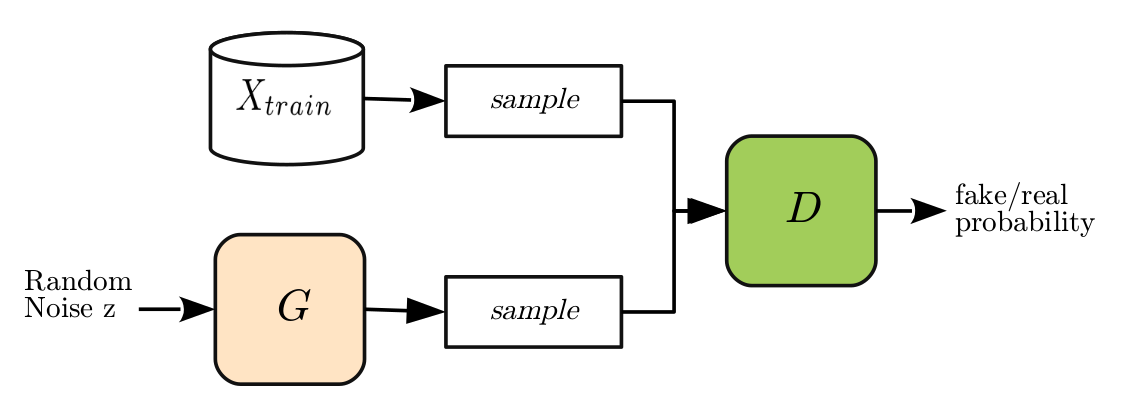}
        \caption{Generative Adversarial Network (GAN).}
        \label{fig:gan}
        \vspace{-0.3cm}
\end{figure}

\descr{Generative Adversarial Networks (GANs)~\cite{goodfellow2014generative}} are neural networks trained in an adversarial manner to generate data mimicking some distribution.
The main intuition is to have two competing neural network models. One takes noise as input and generates samples--and so is called the \emph{generator}. The other model, the \emph{discriminator}, receives samples from both the generator and the training data, and has to be able to distinguish between the two sources. 
The two networks play a continuous game where the generator is learning to produce more and more realistic samples, and the discriminator is learning to get better and better at distinguishing generated data from real data, as depicted in Fig.~\ref{fig:gan}.

More formally, to learn the generator's output distribution over data-points $\mathbf{x}$, we define a prior on input noise variables $p_\mathbf{z} ( \mathbf{z} ) $, then represent a mapping to data space as $G( \mathbf{z} ; \theta_g )$, where $G$ is a generative deep neural network with parameters $\theta_g$.
We also define a discriminator $D ( \mathbf{x} ; \theta_d )$ that outputs $D ( \mathbf{x} ) \in [0,1]$, representing the probability that $\mathbf{x}$ was taken from the training set rather than from the generator $G$. 
$D$ is trained to maximize the probability of assigning the correct label to both real training examples and fake samples from $G$. 
We simultaneously train $G$ to minimize $\log(1 - D ( G ( \mathbf{z} )))$.
The final optimization problem solved by the two networks $D$ and $G$ follows a two-player minimax game as: \vspace*{-0.25cm}
\begin{equation*}
\resizebox{1\columnwidth}{!} 
{
  $\min_{G} \max_{D} \: \mathbb{E}_{\mathbf{x} \sim p_{data}( \mathbf{x} )} [\log D ( \mathbf{x} )] + \mathbb{E}_{\mathbf{z} \sim p_{\mathbf{z}}( \mathbf{z} )}[\log(1 - D ( G ( \mathbf{z} )))]\vspace*{-0.25cm}$ 
  }
\end{equation*}

\noindent First, gradients of $D$ are computed to discriminate fake samples from training data, then $G$ is updated to generate samples that are more likely to be classified as data.
After several steps of training, if $G$ and $D$ have enough capacity and a Nash equilibrium is achieved, they will reach a point at which both cannot improve~\cite{goodfellow2014generative}.

Recently, Lucic et al.~\cite{lucic2017gans} show that, despite a large number of proposed changes to the original GAN model~\cite{berthelot2017began, arjovsky2017wasserstein, gulrajani2017improved} it is still difficult to assess if one performs better than another. 
They also show that the original GAN performs equally well against other state-of-the-art GANs, concluding that any improvements are due to computational budgets and hyper-parameter tuning, rather than scientific breakthroughs.

\descr{Variational Auto-encoders (VAE)~\cite{kingma2013auto}.} %
VAEs~\cite{kingma2013auto} consist of two neural networks (an \emph{encoder} and a \emph{decoder}) and a loss function.
The encoder compresses data into a latent space $(z)$ while the decoder reconstructs the data given the hidden representation.
Rather than attempting to maximize the likelihood, one could maximize a lower bound of the likelihood, thus, if the lower bound increases to a given level, the likelihood must be at least as high.
If hidden variables are continuous, the lower bound, introduced by Variational Auto-encoders (VAEs), can be used.
More formally, let $\mathbf{x}$ be a random vector of observed variables, which are either discrete or continuous. 
Let $\mathbf{z}$ be a random vector of latent continuous variables.

The probability distribution between $\mathbf{x}$ and $\mathbf{z}$ assumes the form $p_\theta(\mathbf{x}, \mathbf{z}) = p_\theta(\mathbf{z}) p_\theta(\mathbf{x} \mid \mathbf{z})$, where $\theta$ indicates that $p$ is parametrized by $\theta$.
Also, let $q_\phi(\mathbf{z} \mid \mathbf{x})$ be a recognition model whose goal is to approximate the true and intractable posterior distribution $p_\theta(\mathbf{z} \mid \mathbf{x})$.
We can then define a lower-bound on the log-likelihood of $\mathbf{x}$ as follows:
$ \mathcal{L}(\mathbf{x}) = - D_{KL}(q_\phi(\mathbf{z} \mid \mathbf{x}) \mid\mid p_\theta(\mathbf{z}))
  + \mathrm{E}_{q_\phi(\mathbf{z} \mid \mathbf{x})} [\log p_\theta(\mathbf{x} \mid \mathbf{z})]$.
The first term pushes $q_\phi(\mathbf{z} \mid \mathbf{x})$ to be similar to $p_\theta(\mathbf{z})$ ensuring that, while training, the VAE learns a decoder that, at generation time, will be able to invert samples from the prior distribution such they look just like the training data.
The second term can be seen as a form of reconstruction cost, and needs to be approximated by sampling from $q_\phi(\mathbf{z} \mid \mathbf{x})$.
In VAEs, the gradient signal is propagated through the sampling process and through $q_\phi(\mathbf{z} \mid \mathbf{x})$, using the so-called re-parametrization trick.
This is done by making $\mathbf{z}$ be a deterministic function of $\phi$ and some noise $\mathbf{\epsilon}$, i.e., 
$\mathbf{z} = f(\phi, \mathbf{\epsilon})$.
For instance, sampling from a normal distribution can be done as $\mathbf{z}=\mu + \sigma \mathbf{\epsilon}$, where $\mathbf{\epsilon} \sim \mathcal{N}(0, \mathbf{I})$.
The re-parametrization trick can be viewed as an efficient way of \emph{adapting} $q_\phi(\mathbf{z} \mid \mathbf{x})$ to help improve the reconstruction.
VAEs are trained using stochastic gradient descent to optimize the loss w.r.t. the parameters of the encoder and decoder $\theta$ and $\phi$.

Larsen et al.~\cite{larsen2015autoencoding} combine VAEs and GANs into an unsupervised generative model that simultaneously learns to encode and generate new samples, which contain more details, sampled from the training data-points.

\section{Membership Inference Attacks Against Generative Models}\label{sec:outline}

In this section, we present our membership inference attacks against generative models. %

\subsection{Threat Model}\label{ssec:threat_model}

We consider an adversary that aims to infer whether \emph{a single known record} was included in the training set of a generative model. We distinguish between two settings: \emph{black-box} and \emph{white-box}  attacks. %
In the former, the attacker can only make queries to the target model under attack -- which we denote as the \emph{target model} -- and has no access to the internal parameters of the model;
in the latter, they also have access to the parameters of a trained target model.
Overall, the accuracy of the attack is measured as the fraction of the records correctly inferred as members of the training set.

\descr{Assumptions.} In both settings, the adversary knows the size of the training set, but not its original data-points.
Variants of the attack allow the adversary to access some further side information, as discussed below.
In order to evaluate the accuracy of our attacks, we will consider an attacker attempting to distinguish data-points used to train the target model, thus, 
we consider an attacker that has a set with data points they suspect are in the original training records.
However, the construction of the attack does {\em not} depend on access to \emph{any} dataset. 
We also assume the attacker knows the size of the training set,
but does {\em not} know how data-points are split into training and test sets. 

In the white-box attacks, the adversary only needs access to the discriminator of a target GAN model. 
In particular, we consider a setting where target model parameters -- i.e., both generator and discriminator in the target GAN model -- are leaked following a data breach or models initially trained on cloud platforms and then compressed/deployed to mobile devices~\cite{hinton2015distilling}.

\descr{Black-box Setting.} In black-box attacks, we assume the attacker does not have prior or side information about training records or the target model. In particular, the attack proceeds
with \underline{\bf no knowledge} of the following:\smallskip

\begin{enumerate}
    \item {\em Target model parameters and hyper-parameters:} No access to network weights from the trained target model, nor to hyper-parameters such as regularization parameters or number of epochs used to train the target model.\smallskip
    \item {\em Target model architecture:} The attacker has no knowledge of the architecture of the target model.     \smallskip
    \item {\em Dataset used to train the target model:} No knowledge of data-points used to train the target model, or the type of data-points used in training, since this is inferred from sampling the target model at inference 
        time. Note that, by contrast, the membership inference attack on discriminative models by Shokri et al.~\cite{shokri2016membership} {\em does} require some information about the dataset, e.g., the syntactic format of data records used in training, in order to generate synthetic samples used in the attack.
    \item {\em Prediction values:} Shokri et al.~\cite{shokri2016membership}  show that predictions scores leak information used to perform membership inference attacks. However, due to the very nature of generative models, in our attacks, the adversary cannot generate prediction scores directly from the target model.\smallskip
\end{enumerate}

\begin{figure}[t]
        \centering
		\includegraphics[width=.18\textwidth]{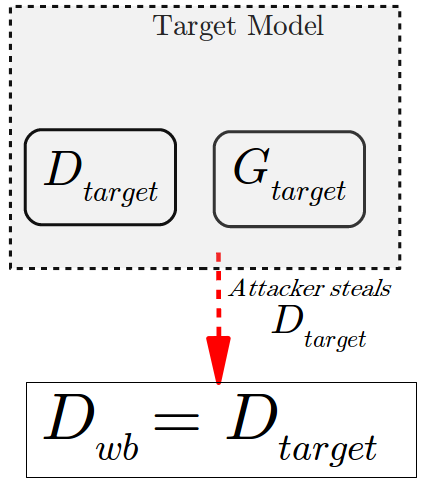}
        \caption{High-level Outline of the White-Box Attack.}
        \label{fig:white_attack}
	\vspace{-0.15cm}   
\end{figure}

\subsection{White-Box Attack}\label{ssec:wb-attack-outline}

We now present our white-box attack; a high-level description is given in Fig.~\ref{fig:white_attack}.

To evaluate the attack, here we assume that an attacker \WB has access to 
the trained target model, namely, a GAN -- i.e., a generator $G_{target}$ and a discriminator $D_{target}$. 
The attacker has a dataset, $X=\{x_1,\dots,x_{m+n}\}$,
which they suspect contains data-points used to train the target model, where $n$ is the size of the training set, and $m$ is the number of data-points that do not belong to the training set.

\begin{figure}[t]
   \centering
   \includegraphics[width=0.5\textwidth]{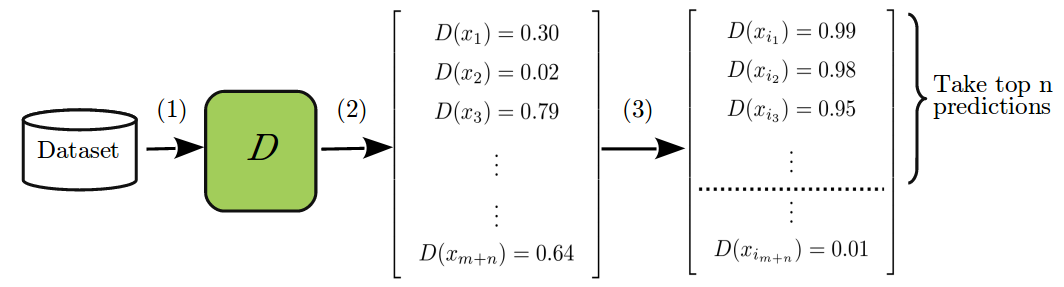}
   \caption{White-Box Prediction Method: The attacker inputs data-points to the Discriminator $D$ (1), extracts the output probabilities (2), and sorts them (3).}
   \label{fig:predictions-fig}
   \vspace{-0.1cm}   
\end{figure}

The target model has been trained to generate samples
that resemble the training set samples. \WB creates a local copy of $D_{target}$, which we refer to as $D_{wb}$.
Then, as shown in Fig.~\ref{fig:predictions-fig}, \WB inputs all samples $X=\{x_1,\dots,x_{m+n}\}$ into $D_{wb}$, which 
outputs the resulting probability vector $\mathbf{p} = [D_{wb}(x_1), \dots, D_{wb}(x_{m+n}) ]$. If the target model overfitted on the training data,
$D_{wb}$ will place a higher confidence value on samples that were part of the training set.
\WB sorts their predictions, $\mathbf{p}$, in descending order and takes the samples associated with the largest $n$ probabilities as predictions for members of the training set. 

Note that the attacker does not need to train a model; rather, it relies 
on internal access to the target model, from which the attack can be launched.

\begin{figure*}[t]
   \centering
    \begin{subfigure}[b]{0.29\textwidth}
        \centering
        \includegraphics[width=\textwidth]{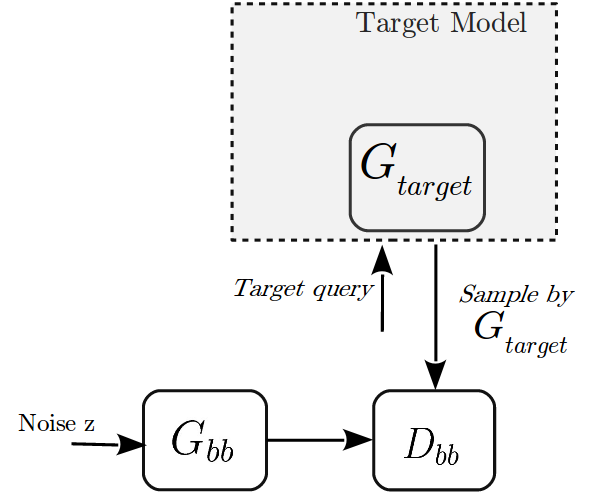}
       \caption{}
       \label{fig:black_attack}
    \end{subfigure}
	~
    \begin{subfigure}[b]{0.32\textwidth}
        \centering
        \includegraphics[width=0.9\textwidth]{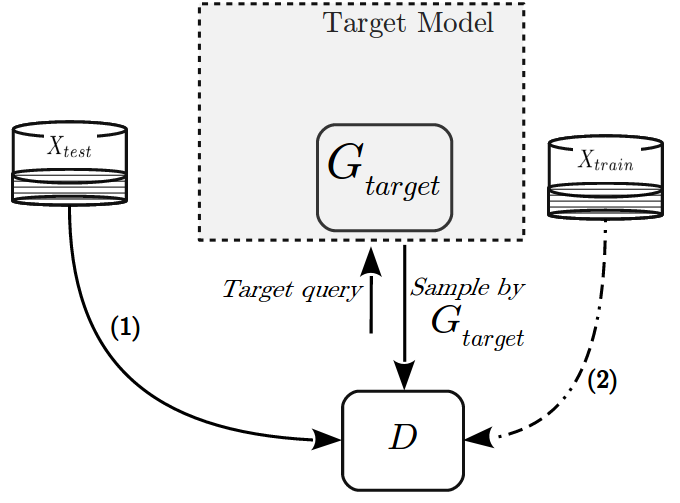}
        \caption{}
        \label{fig:adv_discriminator}
    \end{subfigure}
	~~
    \begin{subfigure}[b]{0.31\textwidth}
      \centering
      \includegraphics[width=\textwidth]{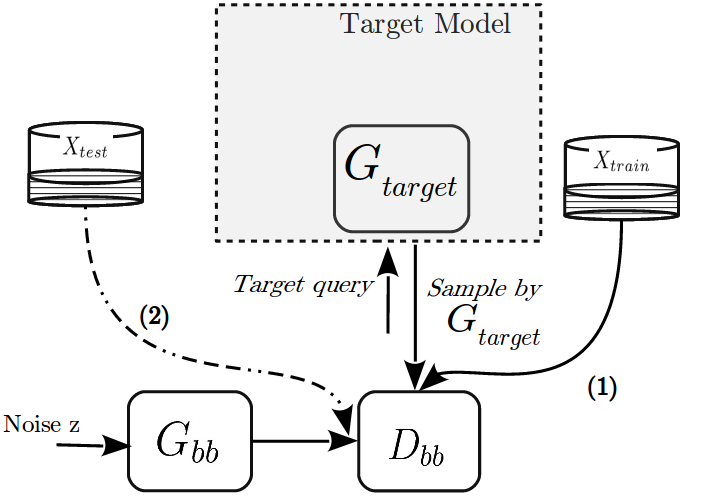}
      \caption{}
      \label{fig:adv_bb}
    \end{subfigure}
    \vspace{-0.3cm}
    \caption{High-level overview of the (a) black-box attack with no auxiliary knowledge, and (b) {\em Discriminative} and (c) {\em Generative} black-box attack with limited auxiliary attacker knowledge.}
    \label{fig:bb-outline}    
    \vspace{-0.2cm}    
\end{figure*}

\subsection{Black-Box Attack with No Auxiliary Knowledge}\label{ssec:bb-attack-outline-1}

In the black-box setting, we assume that the attacker \BB does not have access to the target model parameters. Thus,
\BB cannot directly steal the discriminator model
from the target as in the white-box attack. Furthermore, while in the white-box attack we
restrict the target model to be a GAN, here we do not, and the target model may
not have an associated discriminative model (as with VAEs). 

Again, to evaluate the attack, we assume the attacker has a dataset, $X=\{x_1,\dots,x_{m+n}\}$,
with data-points suspected to have been used to train the target model, where $n$ is the size of the training set. However, the attacker has no knowledge of 
how the training set was constructed from $X$, thus, they do no have access to the true
labels of samples from the dataset and so cannot train a model using a discriminative approach. Instead, \BB trains a GAN in order to re-create the target model locally and, in the process, creates a discriminator $D_{bb}$, which detects overfitting in the generative target model $G_{target}$. 

We illustrate the attack in Fig.~\ref{fig:black_attack}.
Specifically, 
\BB locally trains a GAN ($G_{bb}, \:D_{bb}$) using queries from the target, i.e., \BB trains the local GAN on samples generated by $G_{target}$. As the black-box
attack depends only on samples generated by the target model, $G_{target}$ can be any generative model. 
We assume \BB has neither knowledge nor control over the source of randomness used to generate the samples generated by $G_{bb}$.  
After the GAN has been trained, the attack proceeds to the white-box setting, i.e.,
\BB inputs data-points $X$ into $D_{bb}$, sorts the resulting probabilities, and takes the largest $n$ points as predictions for the training set (as shown in Fig.~\ref{fig:predictions-fig}). 

\subsection{Black-Box Attack with Limited Auxiliary Knowledge}
\label{ssec:bb-attack-outline-2}

In the black-box attack presented above, we assume that \BB has no additional knowledge about subsets of members of the dataset. 
However, we also study the case where an attacker could leverage limited additional side information about the training set. This is a realistic setting, which has been considered extensively in the literature; for instance, social graph knowledge has been used to de-anonymize social networks~\cite{narayanan2009anonymizing}. Overall, auxiliary/incomplete knowledge of sensitive datasets is a common assumption in literature~\cite{qian2016anonymizing, ji2015your}.
Further, the attacker might be able to collect additional information, e.g., from pictures on online social networks or from datasets leaked from data breaches, where the pictures have been used to train the target model under attack.
Access to side information about the training set means that the attacker can ``augment'' the black-box attack.
We consider two settings: a generative and a discriminative one; %
in either, the attacker has incomplete knowledge of members of the test dataset, the training dataset, or both.

\descr{Discriminative setting.} We consider an attacker
that trains a simple discriminative model to infer membership of the training set, as illustrated in Fig.~\ref{fig:adv_discriminator}.
This is feasible since the attacker now has access to membership binary labels, i.e., whether data points belong to the training set or not.
Thus, they do not need to train a generative model to detect overfitting.
Within this setting, we consider two scenarios where the attacker has limited auxiliary knowledge of:\smallskip
\begin{enumerate}
    \item[(1)] Samples that \emph{were not}
used to train the target model; 
    \item[(2)] \emph{Both} training set and test set samples.\smallskip
\end{enumerate}
In both cases, the general method of attack is the same: an attacker trains a local model to detect overfitting in the target model.
In (1), the discriminator, $D$, is fed samples from this auxiliary set, labeled as fake samples, and samples generated by the target model, 
labeled as real samples. If the target model overfits the training set, $D$ will learn to discriminate between training and test samples. 
In (2), $D$ is fed both target generated samples
and the auxiliary training samples, labeled as real samples, and samples from the auxiliary test set, labeled as fake.
Once the attacker has trained a discriminator,
the attack again proceeds as described in Fig.~\ref{fig:predictions-fig}.
Note that we have to consider that the attacker knows some test samples (i.e., fake samples) in order to properly train a binary discriminator.

\descr{Generative setting.} We also consider a generative attack, 
as outlined in Fig.~\ref{fig:adv_bb},
again, as per two scenarios, where the attacker has limited auxiliary knowledge of:\smallskip
\begin{enumerate}
    \item[(1)] Samples that \emph{were} used to train the target model;
    \item[(2)] \emph{Both} training set and test set samples.\smallskip
\end{enumerate}
With both, the attacker trains a local model---specifically, a GAN---that aims to detect overfitting in the target model.
In (1), the discriminator of the 
attacker GAN, $D_{bb}$, is trained using samples generated by $G_{bb}$, labeled as fake samples, and both samples from the auxiliary training set and target generated samples, labeled
as real. Intuitively, we expect the attacker model to be stronger at recognizing overfitting in the target model, if it has auxiliary knowledge of samples on which it
was originally trained. In (2), $D_{bb}$ is trained on samples generated
by $G_{bb}$ and samples from auxiliary set of test ones, labeled as fake samples, and samples generated by the target model and 
samples from the auxiliary training set, labeled as real.
The attacker GAN is trained to learn to discriminate between test and training samples directly. Again, once the attacker has trained their model, data-points from $X$ are fed into $D_{bb}$, and their predictions are sorted as per Fig.~\ref{fig:predictions-fig}.

\section{Evaluation}\label{sec:experiments}

In this section, we present an experimental evaluation of the attacks described above.

\subsection{Experimental Setup}\label{ssec:experimental}

\noindent\textbf{Testbed.} Experiments are performed using PyTorch %
on a workstation running Ubuntu Server 16.04 LTS, equipped with a 3.4GHz CPU i7-6800K, 32GB RAM, and an NVIDIA Titan X GPU card. 
Source code is available upon request and will be made public along with the final version of the paper.

\descr{Settings.} For white-box attacks, we measure membership inference accuracy at successive epochs of training the target model, where one epoch corresponds to one round of training on all training set inputs.\footnote{We update model weights after training on mini-batches of 32 samples.} 
For black-box attacks, we fix the target model and measure membership
inference accuracy at successive training steps of the attacker model, where one training step is defined as one iteration of training on a mini-batch of inputs.
The attacker model is trained using soft and noisy labels as suggested in~\cite{salimans2016}, i.e., we replace labels with random numbers in $[0.7,1.2]$ for real samples, and random values in $[0.0, 0.3]$ for fake samples. Also, we occasionally flip the labels when training the discriminator.
These GAN modifications are known to stabilize training in practice~\cite{ganhacks}. 

\descr{Datasets.} We perform experiments using two popular image datasets as well as a health-related dataset:\smallskip
\begin{enumerate}
\item  Labeled Faces in the Wild (LFW)~\cite{LFWTech}, which includes 13,233 images of faces collected from the Web;
\item CIFAR-10~\cite{krizhevsky2009learning}, with 60,000 32x32 color images in 10 classes, with 6,000 images per class;
\item Diabetic Retinopathy (DR)~\cite{dr}, consisting of 88,702 high-resolution retina images taken under a variety of image conditions.\smallskip
\end{enumerate}

For LFW and CIFAR-10, we randomly choose 10\% of the records as the training set. 
The LFW dataset is ``unbalanced,'' i.e., some people appear in multiple images, while others only appear once.
We also perform experiments so that the training set is chosen to include the ten most popular classes of people in terms of number of images they appear in, which amounts to 12.2\% of the LFW dataset. 
Intuitively, we expect that models trained on the top ten classes will overfit more than the same models trained on random 10\% subsets, as we are training on a more homogeneous set of images.

Note that experiments using the DR dataset are presented in Section~\ref{sec:eval_eye}, which discusses a case-study evaluation on a dataset of medical relevance.
From DR, we select images with moderate to proliferate diabetic retinopathy presence, and use them to train the generative target model.

\begin{figure*}[t]
   \centering
   \begin{subfigure}[b]{\suba\textwidth}
        \centering
        \includegraphics[width=\textwidth]{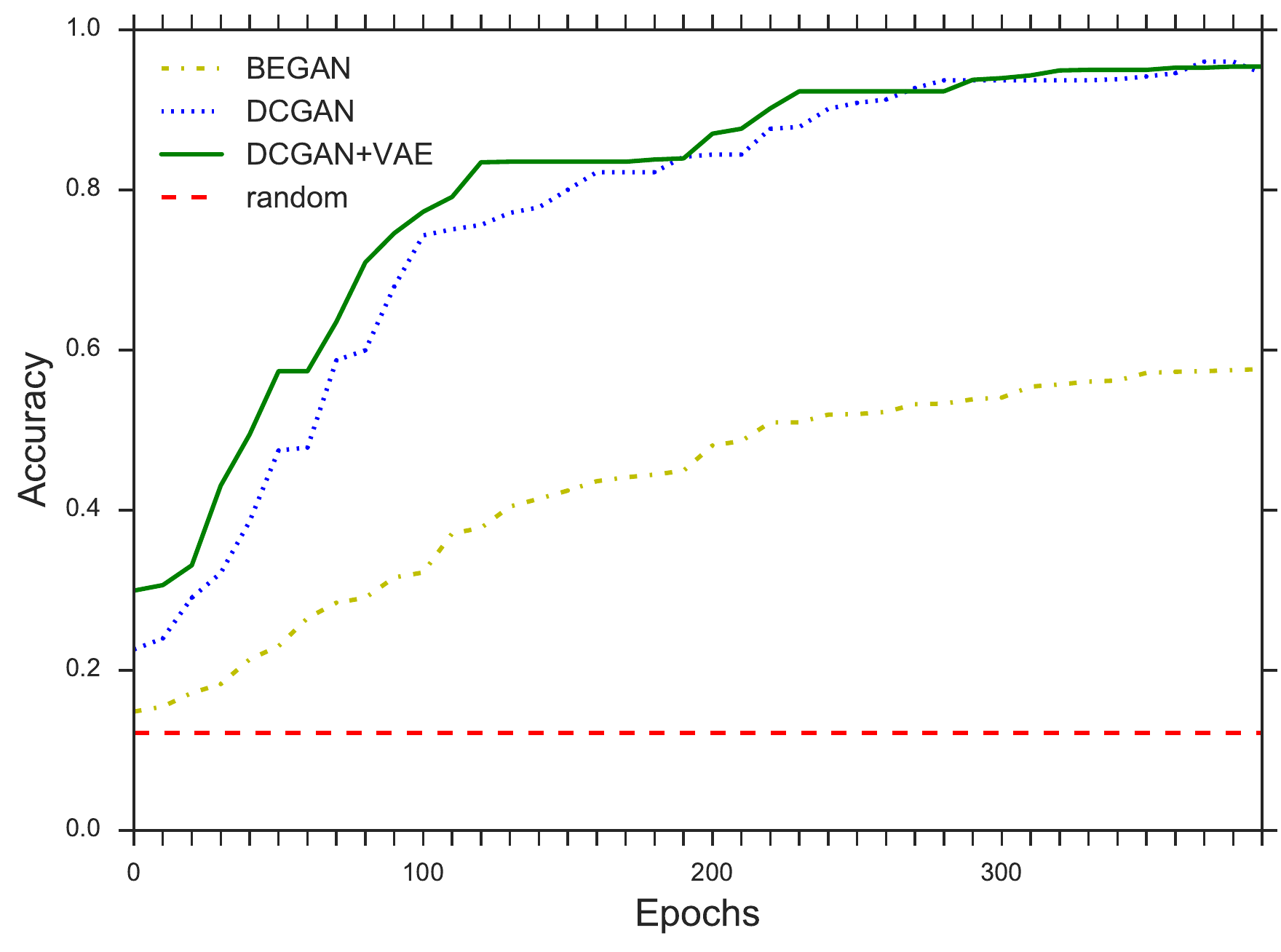}
        \caption{LFW, top ten classes}
        \label{fig:smoothed-wb-lfw-top10}
    \end{subfigure}
    ~
     \begin{subfigure}[b]{\suba\textwidth}
        \centering
        \includegraphics[width=\textwidth]{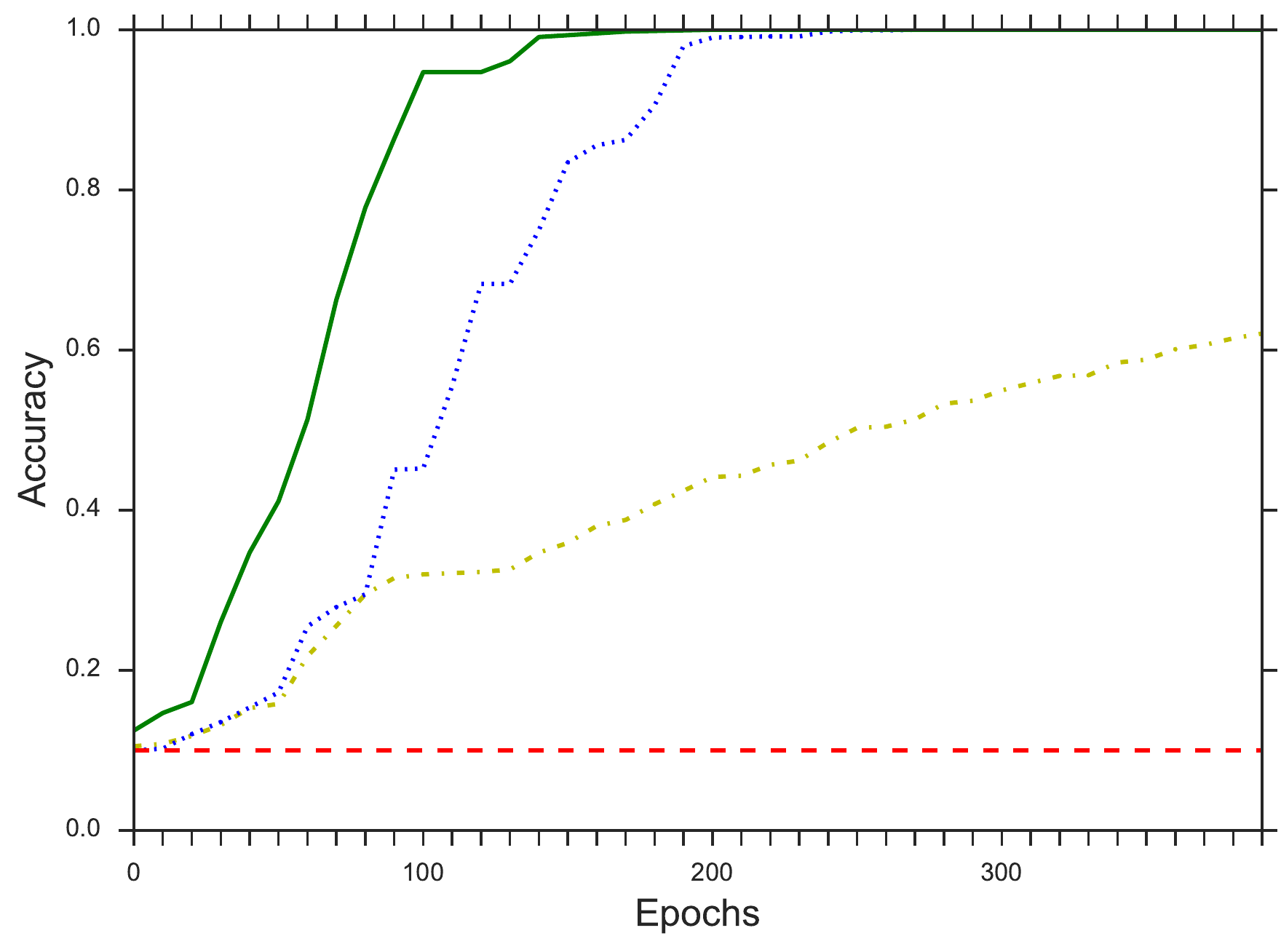}
        \caption{LFW, random 10\% subset}
        \label{fig:smoothed-wb-lfw-rand10}
    \end{subfigure}
    ~
    \begin{subfigure}[b]{\suba\textwidth}
        \centering
        \includegraphics[width=\textwidth]{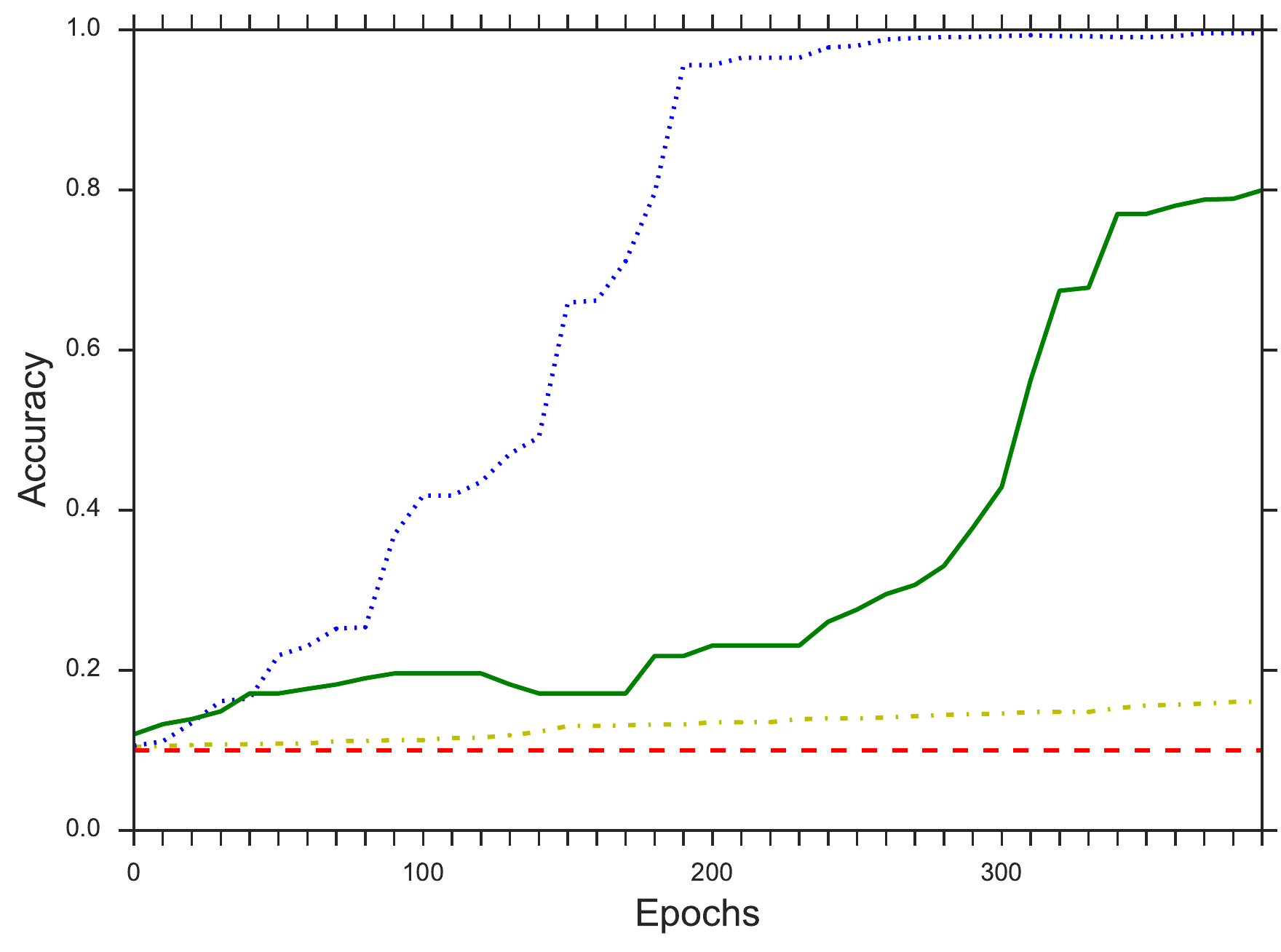}
        \caption{CIFAR-10, random 10\% subset}
        \label{fig:smoothed-wb-cifar-rand10}
    \end{subfigure}
    \vspace{-0.6cm}    
    \caption{Accuracy of white-box attack with different datasets and training sets.}
    \label{fig:wb-results}
    \vspace{-0.2cm}
\end{figure*}

\descr{Models.} 
Since their introduction, a few GAN~\cite{goodfellow2014generative} variants have been proposed to improve training stability and sample quality.  
In particular, deep convolutional generative adversarial networks (DCGANs)~\cite{radford2015unsupervised} combine the GAN training process
with convolutional neural networks (CNNs). CNNs are considered the state of the art for a range image recognition tasks; by combining CNNs
with the GAN training processes, DCGANs perform well at unsupervised learning tasks such as generating complex 
representations of objects and faces~\cite{radford2015unsupervised}. 
GANs have also been combined with VAEs~\cite{larsen2015autoencoding}: by collapsing the generator (of the GAN) and
decoder (of the VAE) into one, the model uses learned feature representations in the GAN discriminator as the reconstructive error term in the VAE. 
It has also been shown that combining the DCGAN architecture with a VAE yields more realistic generated samples~\cite{otoro}.
More recently, Boundary Equilibrium GAN (BEGAN)~\cite{berthelot2017began} have been proposed as an approximate measure of convergence. Loss terms in GAN 
training do not correlate with sample quality, making it difficult for a practitioner to decide when to stop training. This decision is usually
performed by visually inspecting generated samples. BEGAN proposes a new method for training GANs by changing the loss function.
The discriminator is an autoencoder and the loss is a function of the quality of reconstruction achieved by the discriminator on both generated and real 
samples. BEGAN produces realistic samples~\cite{berthelot2017began}, and is simpler to train since loss convergence and sample quality
is linked with one another.

We evaluate our attacks using, as the target model:\smallskip
\begin{enumerate}
\item DCGAN~\cite{radford2015unsupervised},
\item DCGAN+VAE~\cite{larsen2015autoencoding}, and
\item BEGAN~\cite{berthelot2017began},\smallskip
\end{enumerate}
while fixing DCGAN as the attacker model. 
This choice of models is supported by recent work~\cite{lucic2017gans}, which shows that no other GAN model performs significantly better than our choices.~\cite{lucic2017gans} also demonstrates that VAE models perform significantly worse than any GAN variant. %

\begin{figure*}[t]
   \centering
   \begin{subfigure}[b]{\suba\textwidth}
        \centering
        \includegraphics[width=\textwidth]{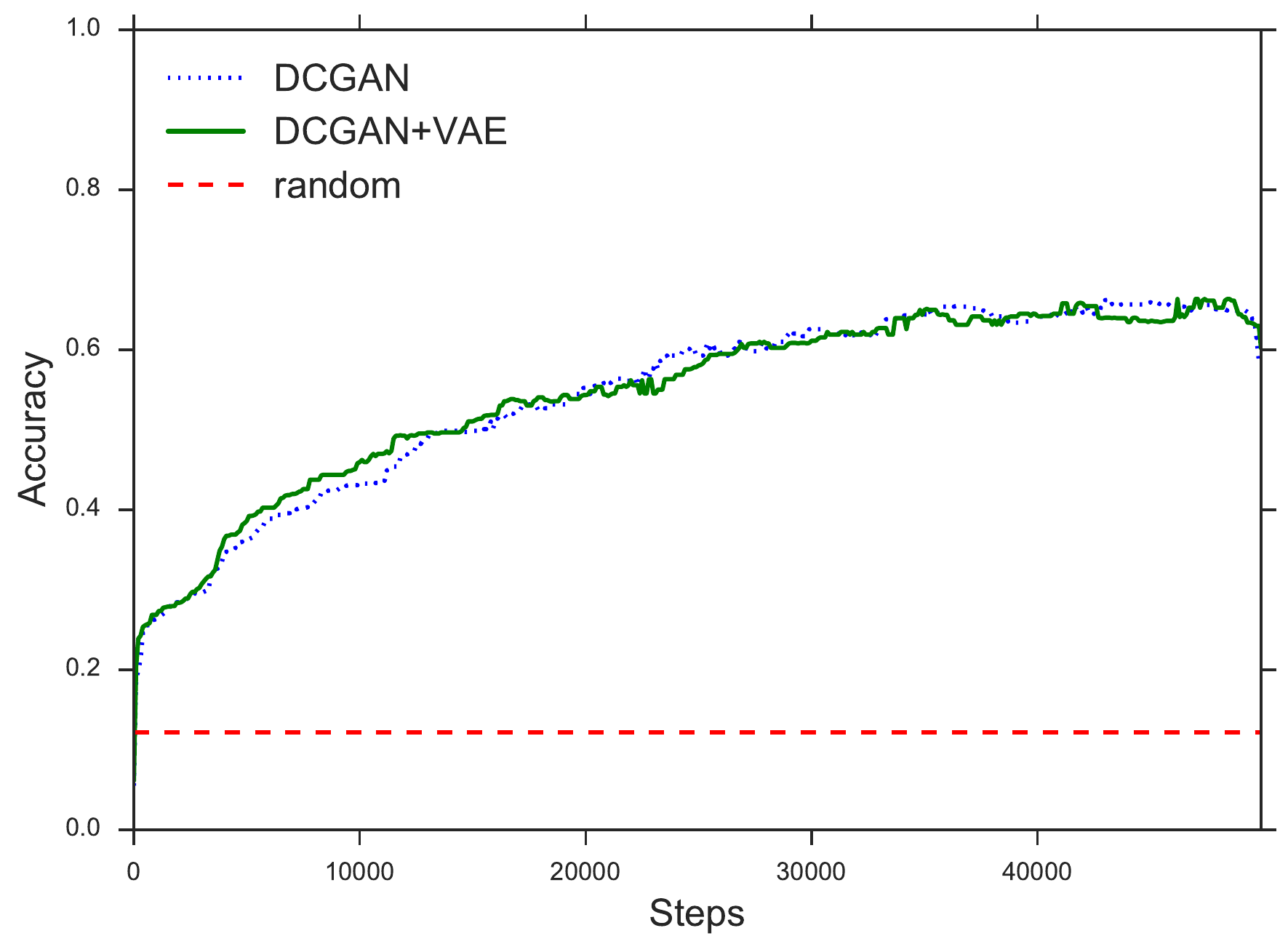}
        \caption{LFW, top ten classes}
        \label{fig:smoothed-bb-lfw-top10}
    \end{subfigure}
    ~
    \begin{subfigure}[b]{\suba\textwidth}
        \centering
        \includegraphics[width=\textwidth]{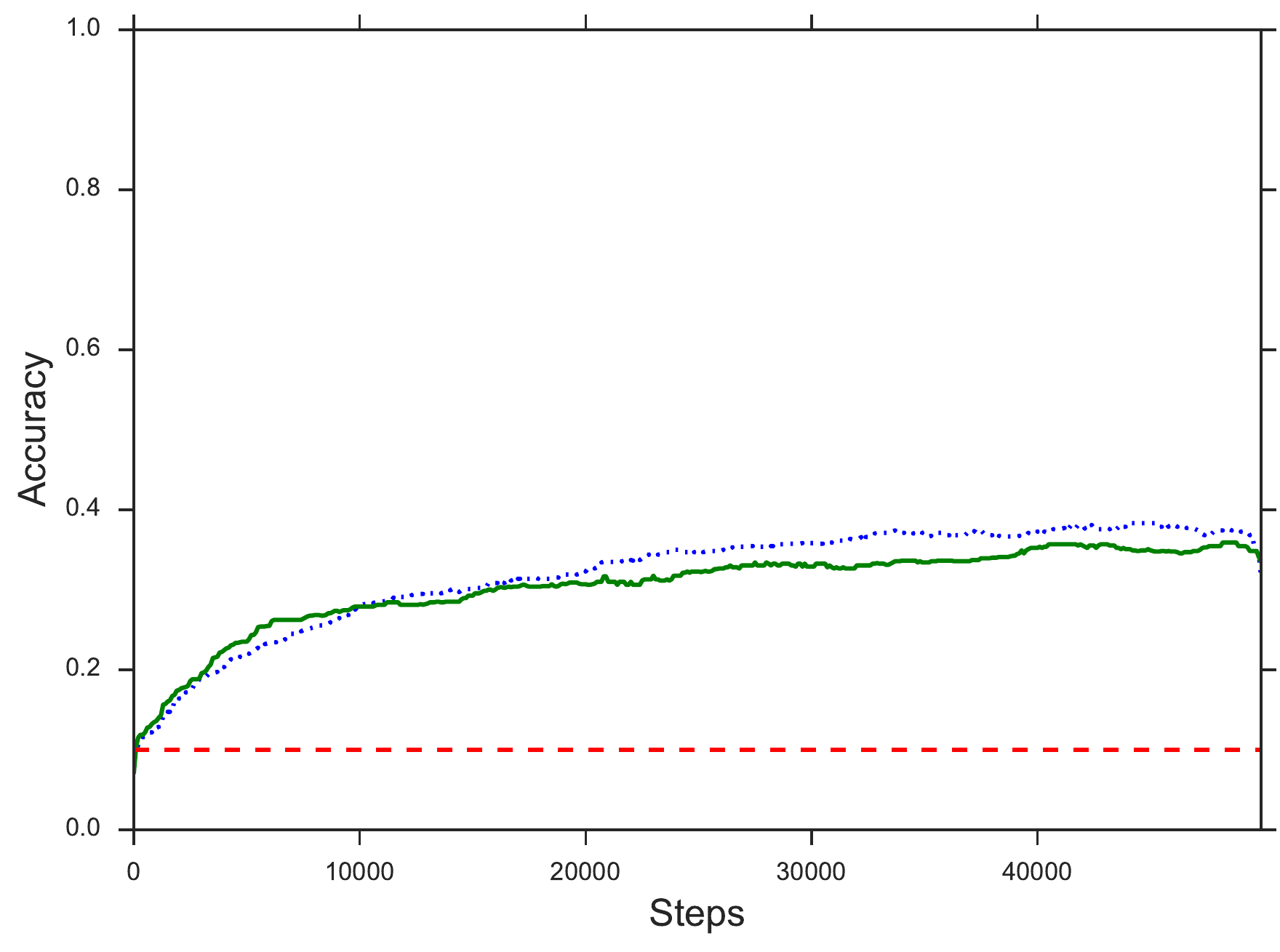}
        \caption{LFW, random 10\% subset}
        \label{fig:smoothed-bb-lfw-rand10}
    \end{subfigure}
    ~
    \begin{subfigure}[b]{\suba\textwidth}
        \centering
        \includegraphics[width=\textwidth]{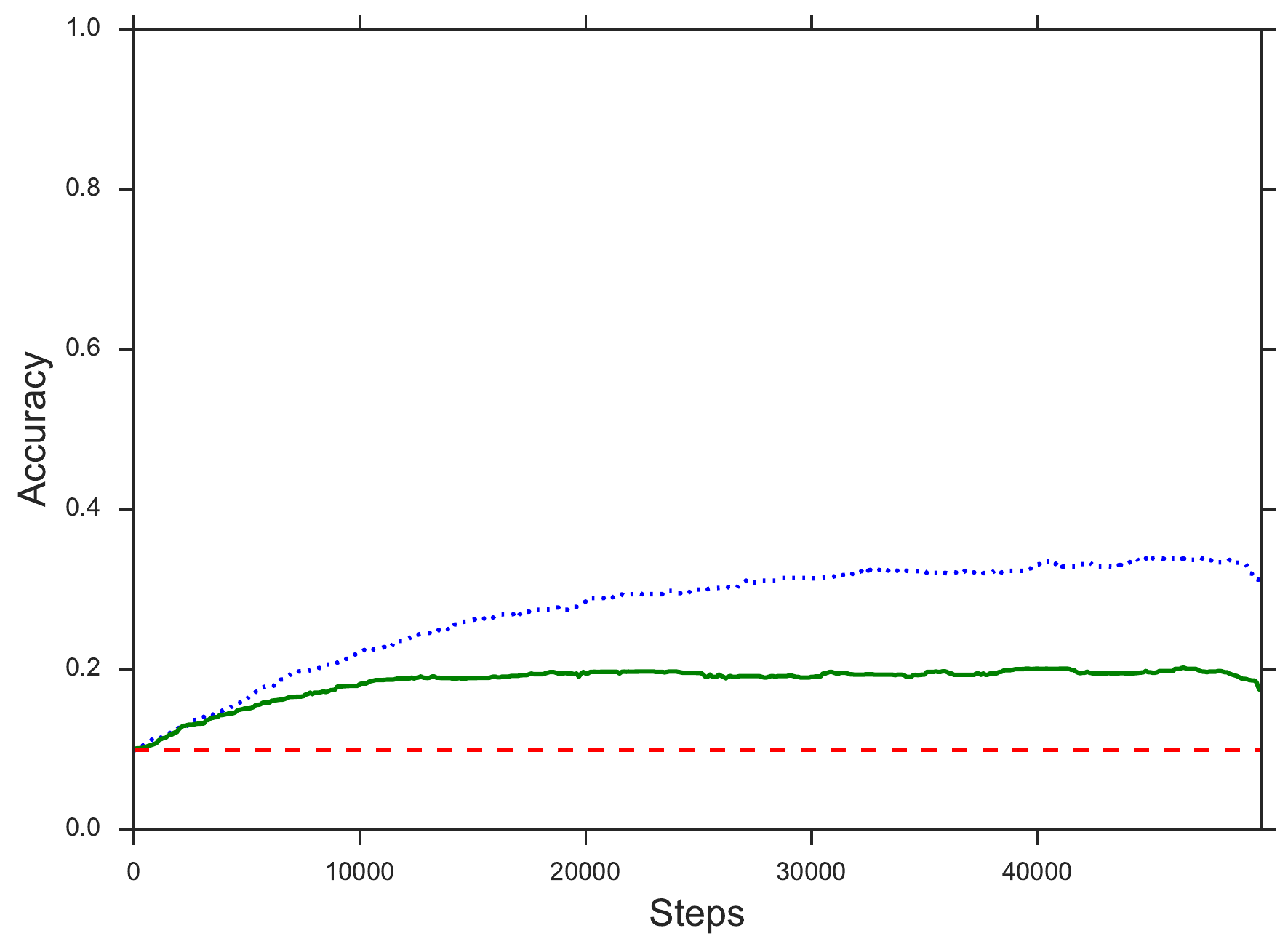}
        \caption{CIFAR-10, random 10\% subset}
        \label{fig:smoothed-bb-cifar-rand10}
    \end{subfigure}
    \vspace{-0.4cm}    
    \caption{Accuracy of black-box attack on different datasets and training sets.}
    \label{fig:bb-results}
    \vspace{-0.2cm}    
\end{figure*}

\subsection{Strawman Approaches}\label{ssec:euclidean}
We begin our evaluation with a na\"ive Euclidean distance based attack.
Given a sample generated by a target model, the attacker computes the Euclidean distance between the generated sample and every real sample in the dataset. 
Repeating this multiple times for newly generated samples, the attacker computes an average distance from each
real sample, sorts the average distances, and takes the smallest $n$ distances (and the associated real samples) as the guess for the training set, where $n$ is the size of the training set.

We perform this attack on a target model (DCGAN) trained on a random 10\% subset of CIFAR-10 and 
a random 10\% subset of LFW, finding that the attack does not perform better than if the 
attacker were to randomly guess which real samples were part of the original training set. 
For completeness, results are reported in Fig.~\ref{fig:euclidean-attack-fig} 
in Appendix~\ref{app:unsuccessful}.
In Appendix~\ref{app:unsuccessful}, we also discuss another unsuccessful approach, based on training a shadow model, inspired by the techniques proposed by Shokri et al.~\cite{shokri2016membership}.

\subsection{White-Box Attack}\label{ssec:wb-exp}
We now present the results of our evaluation of the white-box attack described in Section~\ref{ssec:wb-attack-outline} on LFW and CIFAR-10.
For the LFW dataset, we build the training set either as a random 10\% subset of the dataset or the top ten classes. For CIFAR-10, the training set is a random 10\% subset of the dataset.
The target models we implement are DCGAN, DCGAN+VAE, and BEGAN. 
In the rest of this section, we will include a baseline in the plots (red dotted line) that corresponds to the success of an 
attacker randomly guessing which samples belong to the training set.

Fig.~\ref{fig:smoothed-wb-lfw-top10} shows the accuracy of a white-box attack against a target model trained on the top ten classes of the LFW dataset. 
We observe that both DCGAN and DCGAN+VAE are vulnerable to the white-box attack.
For DCGAN and DCGAN+VAE target models trained for 100 epochs, the attacker 
infers training set membership with 80\% accuracy, and for models trained for 400 epochs --
with 98\% and 97\% accuracy, respectively. The BEGAN target model does overfit, although to a lesser extent: after 400 epochs, an attacker with 
white-box access to the BEGAN target model can infer membership of the training set with 60\% accuracy.
In Fig.~\ref{fig:smoothed-wb-lfw-rand10}, we report the results of white-box attacks against a target model trained on a random 10\% subset of the LFW dataset. 
Similar to Fig.~\ref{fig:smoothed-wb-lfw-top10}, both 
DCGAN and DCGAN+VAE are vulnerable: when these are trained for 250 epochs,
an attacker can achieve perfect training set membership inference. BEGAN performs
similar to the top ten classes white-box experiment, achieving 62\% accuracy after 400 epochs.
Finally, Fig.~\ref{fig:smoothed-wb-cifar-rand10} plots the accuracy of the white-box attack against a target model trained on a random 10\% subset of CIFAR-10. 

For DCGAN, results are
similar to DCGAN on LFW, with perfect training set membership inference after 400 epochs. However, DCGAN+VAE does not leak information (does not overfit)
until around 250 epochs, where
accuracy remains relatively steady, at 10-20\%. Instead, after 250 epochs, the model overfits, with accuracy reaching 80\% by 400 epochs. BEGAN, while 
producing quality samples, does not overfit, with final training set membership inference accuracy of 19\%, i.e., only 9\% better than random guess. Due to the limited accuracy of BEGAN in comparison to other models, we discard it as a target model for black-box attacks as it does not seem to be vulnerable to membership inference attacks.
Note that GAN models need to be trained for hundreds of epochs before reaching good samples quality. 
Indeed, the original DCGAN/BEGAN papers report 2x and 1.5x the number of network updates (when adjusted for training set size) as our white-box attack, to train DCGAN and BEGAN, respectively.

In summary, we conclude that white-box attacks infer the training set with up to perfect accuracy when DCGAN and DCGAN+VAE are the target models. On the other hand, BEGAN is less vulnerable to white-box attacks, with up to 62\% accuracy.

\subsection{Black-Box Attack with No Auxiliary Knowledge}\label{ssec=bb-no-knowledge-experiments}
Next, we present the results of the black-box attacks (see Section~\ref{ssec:bb-attack-outline-1}) on LFW and CIFAR-10.
We assume the attacker has no knowledge of the training or test sets other than
the size of the original training set. Once again, for LFW, the training set is either a random 10\% subset of the dataset or the top ten classes, while, for CIFAR-10, the training set is always a random 10\% subset of the dataset.
The target models we implement are DCGAN and DCGAN+VAE (fixed at epoch 400), and the attacker model uses DCGAN.

Fig.~\ref{fig:smoothed-bb-lfw-top10} plots the results of a black-box attack against a target model trained on the top ten classes of the LFW dataset. After training the attacker model on
target queries, the attack achieves 63\% training set membership inference accuracy for both DCGAN and DCGAN+VAE target models. Surprisingly, the attack
performs equally well when the target model differs from the attack model as when the target and attack model are identical. This highlights the fact that the 
attacker does not need to have knowledge of the target model architecture in order to perform the attack.

In Fig.~\ref{fig:smoothed-bb-lfw-rand10}, the results are with respect to a target model trained on a random 10\% subset of the LFW dataset. Once again, we find that DCGAN and DCGAN+VAE target models are 
equally vulnerable to a black-box attack. An attacker with no auxiliary information of the training set can still expect to perform membership inference with 40\% (38\%) accuracy for the DCGAN (DCGAN+VAE) target model.

Finally, Fig.~\ref{fig:smoothed-bb-cifar-rand10} plots the accuracy of a black-box attack against a target model trained on a 
random 10\% subset of the CIFAR-10 dataset. For the DCGAN+VAE target model, accuracy reaches 20\% after 1,000 training steps and stays flat. For the DCGAN target
model, the attacker can infer training set membership with 37\% accuracy, with accuracy improving steadily throughout the attacker model training process.

We observe that the difference in attack success between the DCGAN and DCGAN+VAE target models with CIFAR-10 and the similar success of the two models
with LFW occur in both white-box and black-box attacks. As expected, the best results are obtained when
the attacker and target model have the same architecture. However, the attack does not overwhelmingly suffer under differing architectures. In fact,
in LFW experiments there is a negligible difference in attack success, and, in the CIFAR-10 black-box experiments, the difference in accuracy is approximately 17\%.

In summary, we conclude that our black-box attacks are less successful, compared to white-box attack, in inferring membership, but perform similarly against different target model architectures.

\subsection{Black-Box Attack with Limited Auxiliary Knowledge}
As discussed in Section~\ref{ssec:bb-attack-outline-2}, we also consider black-box attacks where the attacker has some limited auxiliary knowledge of the dataset, and uses this knowledge to recover the full training set.
We now present the results of these attacks on random 10\% subsets of LFW and CIFAR-10 with DCGAN attacker and target models (fixed at epoch 400).

We consider different scenarios where the attacker has
knowledge of 20--30\% of the training set, 20-30\% of the test set, or both. 
Nonetheless, the total number of samples of which the attacker has knowledge is quite modest. For LFW, 20\% of the random 10\% training set corresponds to 264 out of 1,323 images, 20\% of the test set to 2,382 out of 11,910 images, whereas, for CIFAR-10, 20\% of the random 10\% training set amounts to 1,200 out of 6,000 images, and 20\% of the test set to 10,000 out of 50,000 images.
An attacker with auxiliary information of the training and test set has access to labels, and therefore may not need to train a generative model to perform a membership inference
attack on a generative model.
We also show that, while the attacker can train a discriminative model to perform membership inference, such an approach produces worse results than the generative method.

\descr{Discriminative approach.}
If an attacker has access to true labels within the dataset, they can train a discriminative model on these samples in order to learn to classify training samples correctly. For
both LFW and CIFAR-10 DCGAN target models, trained on a random 10\% subset of the dataset, we consider two settings:\smallskip 

\begin{enumerate}
\item[(i)] the attacker has 20\% knowledge of the {\em test} set; or
\item[(ii)] the attacker has 30\% knowledge of both the training and test set.\smallskip
\end{enumerate}

We use the discriminator from DCGAN as the discriminative model trained by the attacker. 
In (i), we pass test set samples to the discriminator labeled as fake samples, and target generated ones labeled as real. 
In (ii), we pass test set samples to the discriminator labeled as fake ones, and target generated and training set samples labeled as real ones.

\begin{figure}[t]
   \centering
   \includegraphics[width=\subb\textwidth]{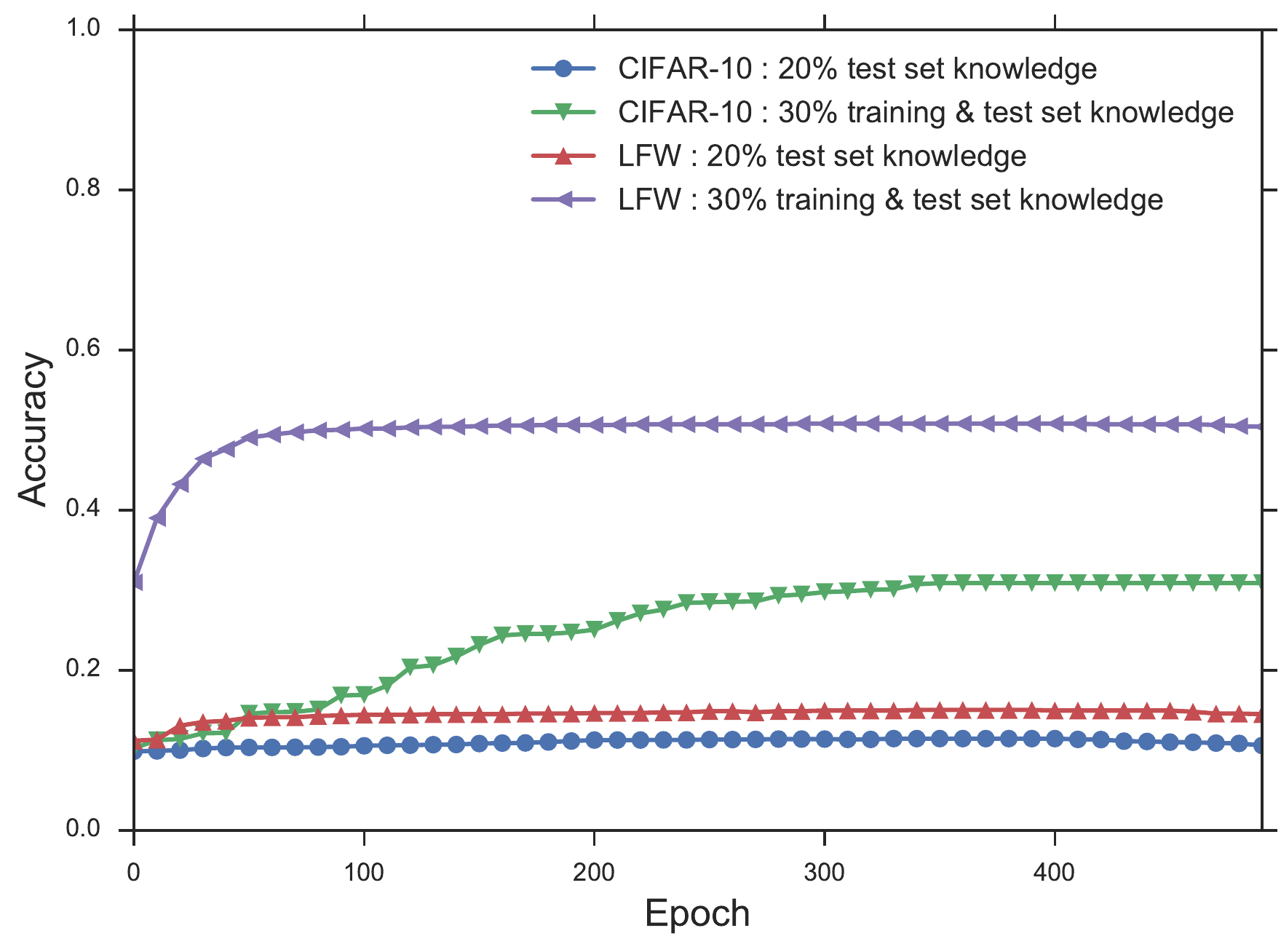}
    \vspace{-0.2cm}
      \caption{Membership inference accuracy using a discriminative model, when the attacker has knowledge of (i) 20\% of the test set, or (ii) 30\% of both training and test sets. Random guess in (i) and (ii) corresponds, respectively, to 14\% and 12\% accuracy.}
   \label{fig:bb-attack-classifier-fig}
\end{figure}

 \begin{figure*}[t]
   \centering
   \begin{subfigure}[b]{\subc\textwidth}
        \centering
        \includegraphics[width=\textwidth]{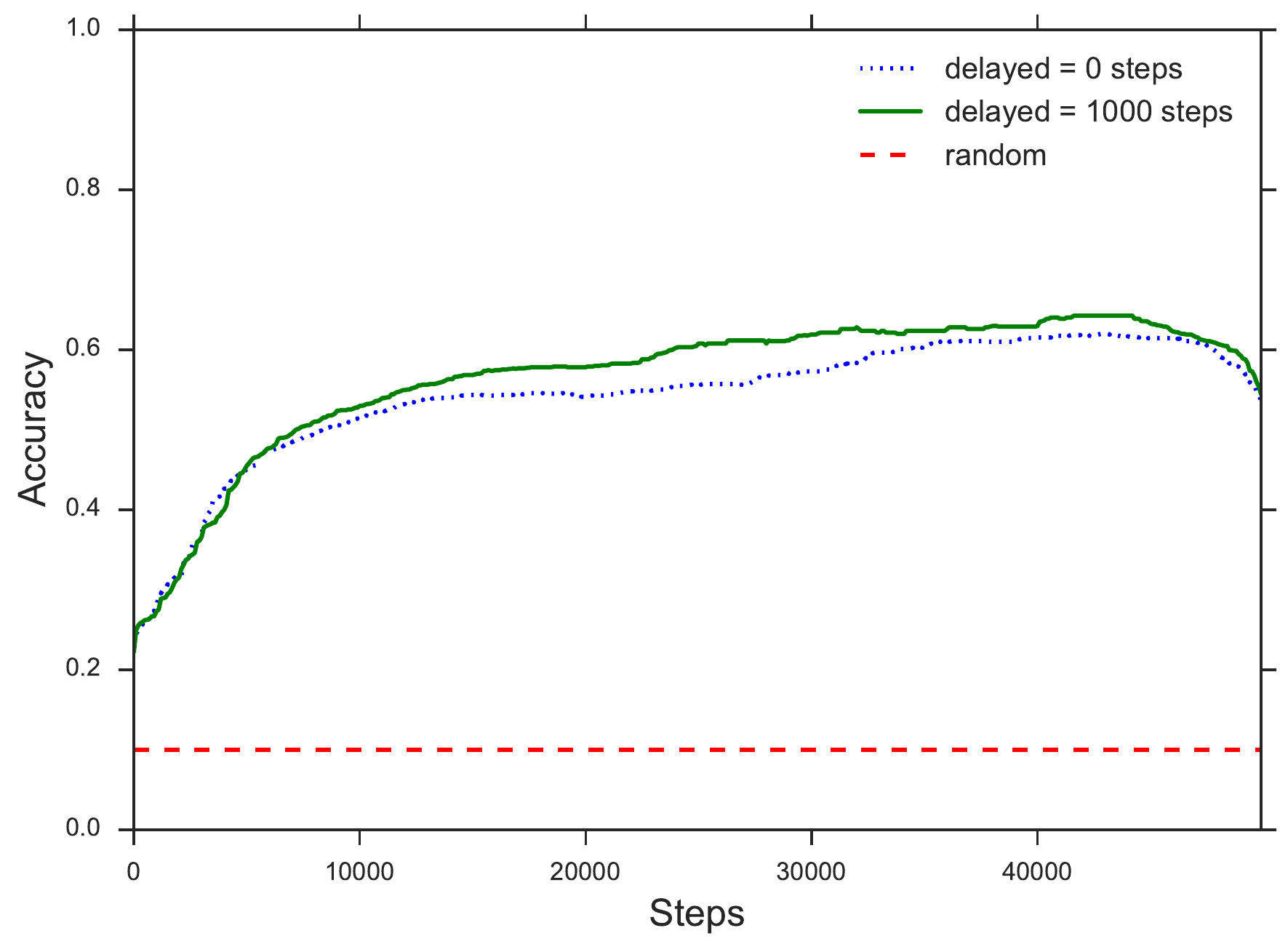}
        \caption{DCGAN}%
        \label{fig:lfw-dcgan-delay-delta}
    \end{subfigure}
    ~~~
	\begin{subfigure}[b]{\subc\textwidth}
        \centering
        \includegraphics[width=\textwidth]{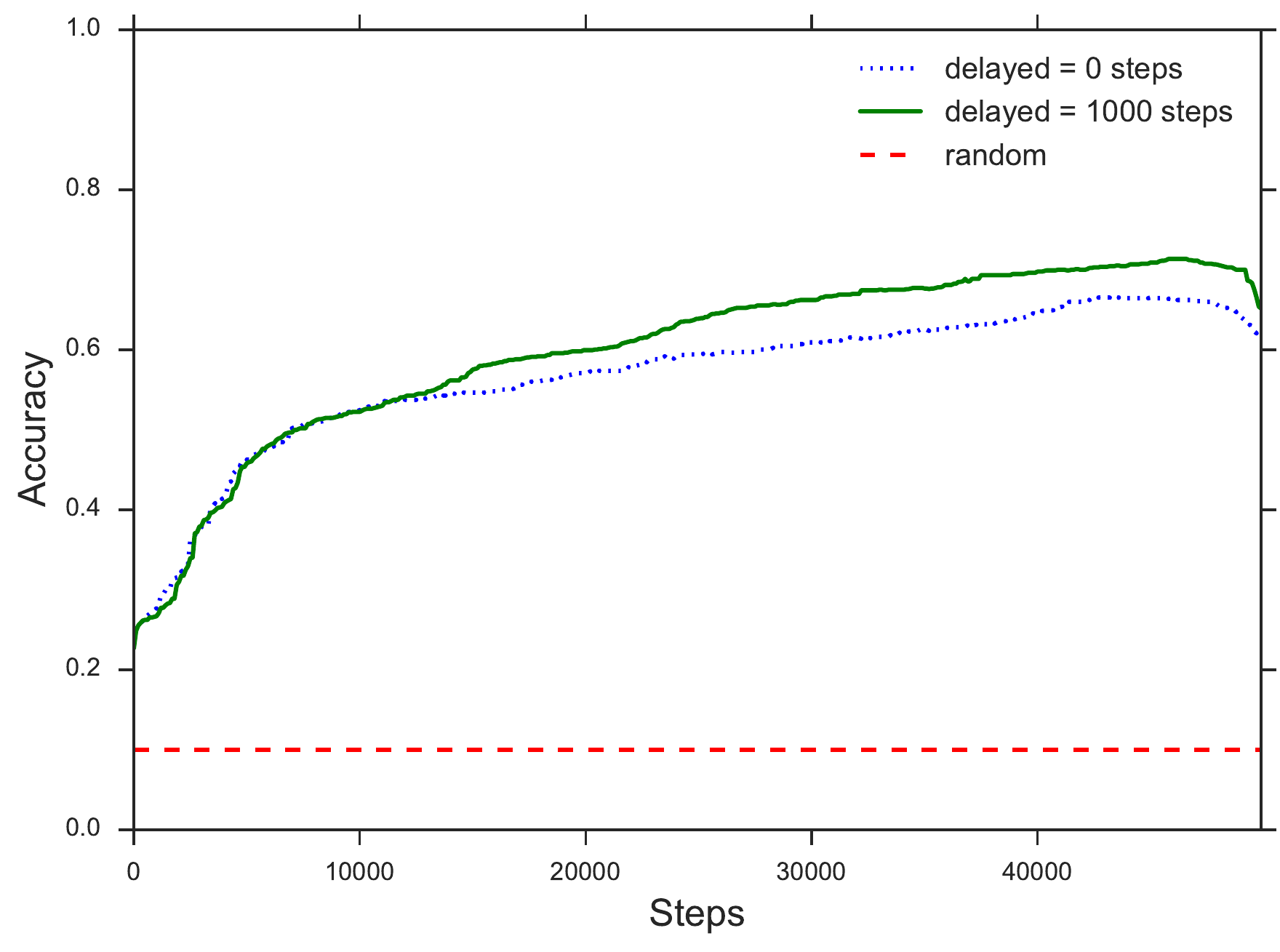}
        \caption{DCGAN+VAE}
        \label{fig:lfw-dcganvae-delay-delta}
    \end{subfigure}
	\vspace{-0.2cm}
    \caption{{Black-box attack results with 20\% attacker training set knowledge for DCGAN/DCGAN+VAE target models, trained on a random 10\% subset of LFW, for different delays at which auxiliary knowledge is introduced into the attacker model training.}}
    \label{fig:lfw-delay-delta}
 \end{figure*}
 
In Fig.~\ref{fig:bb-attack-classifier-fig}, we plot the accuracy results for both settings, 
showing that the attack fails with both datasets when the attacker has only test set knowledge, performing no better than random guessing. Whereas, if the attacker has both training and test knowledge, with LFW, the attacker recovers the training set with 50\% accuracy, while, for CIFAR-10, accuracy reaches 33\%. Note that this approach does not improve
on CIFAR-10 black-box results with no auxiliary knowledge, and only marginally improves on LFW results. As a result, we also experiment with generative approaches to black-box attacks
with auxiliary attacker knowledge, as discussed next.

 \begin{figure*}[t]
   \centering
   \begin{subfigure}[b]{\subc\textwidth}
        \centering
        \includegraphics[width=\textwidth]{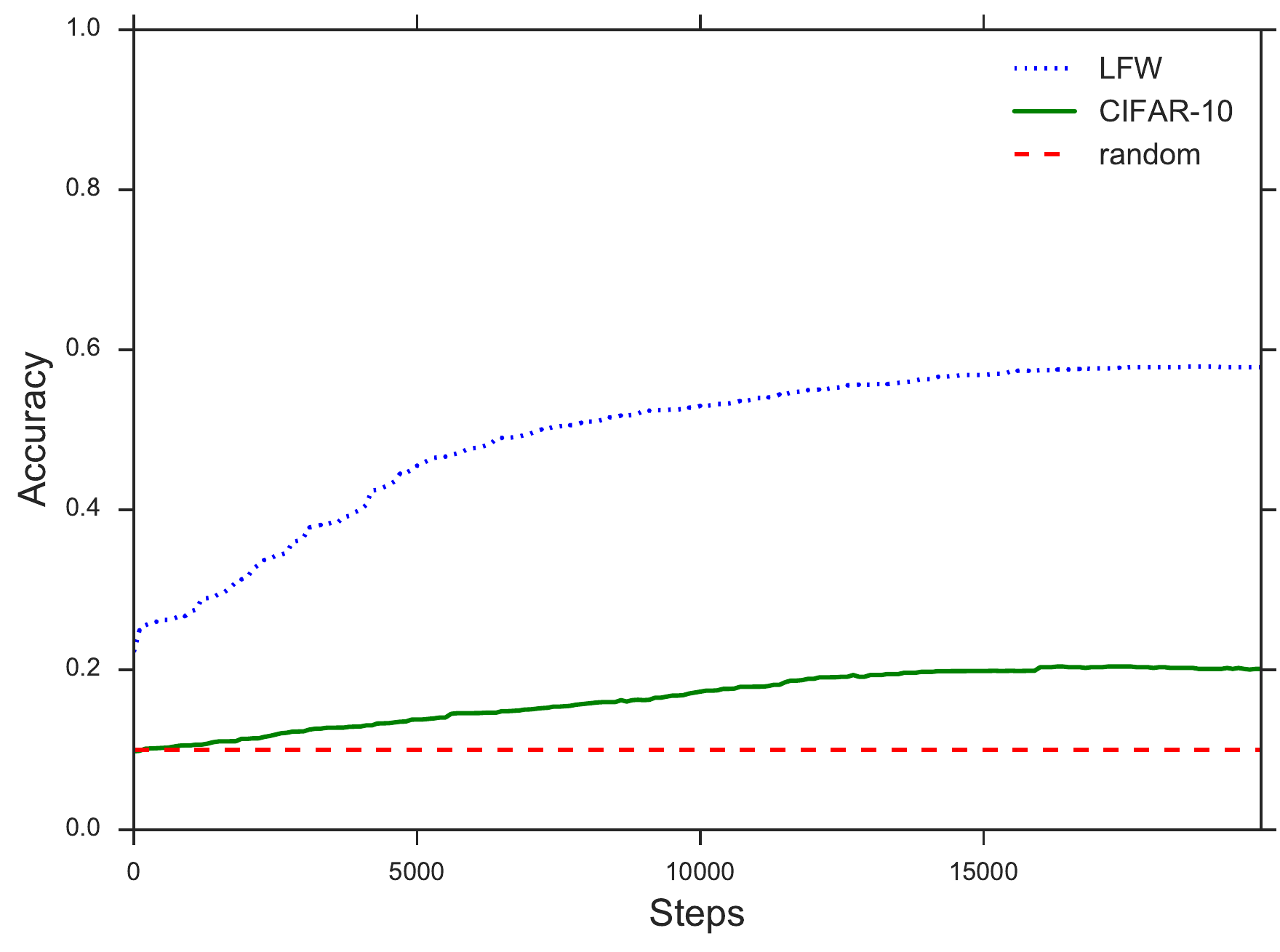}
        \caption{20\% of the training set knowledge}%
        \label{fig:bb-attack-with-adv-train-gen}
    \end{subfigure}
~~~
   \begin{subfigure}[b]{\subc\textwidth}
        \centering
        \includegraphics[width=\textwidth]{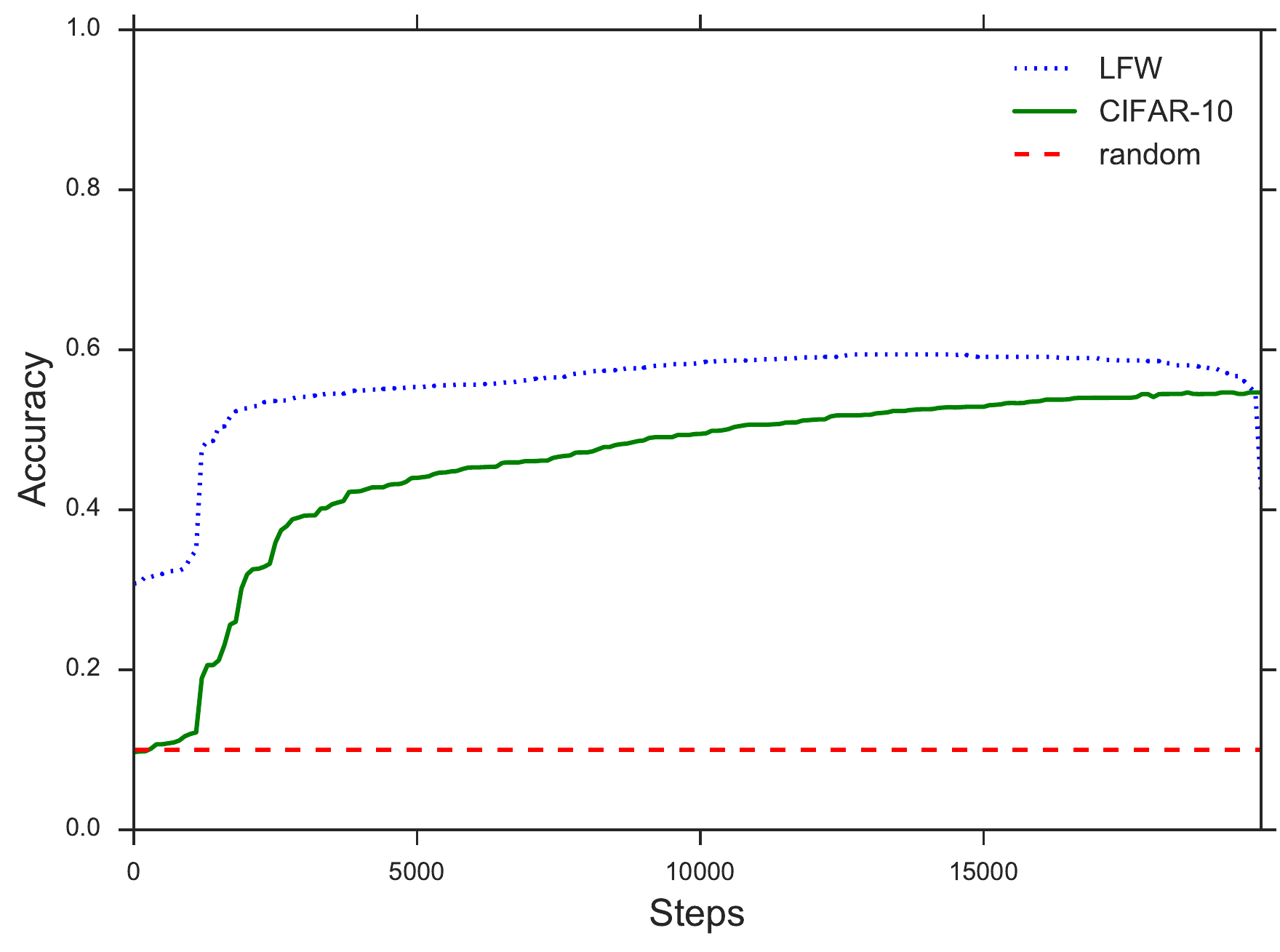}
        \caption{30\% of the training set and test set knowledge}
        \label{fig:bb-attack-with-adv-train-adv-test-gen}
    \end{subfigure}
    \caption{{Black-box results when the attacker has (a) knowledge of 20\% of the training set or (b) 30\% of the training set and test set. 
    The training set is a random 10\% subset of the LFW or CIFAR-10 dataset, and the target model is fixed as DCGAN.}}
    \label{fig:bb-attack-with-adv-know}
 \end{figure*}

\begin{figure*}[t]
   \centering
   \begin{subfigure}[b]{\subb\textwidth}
        \centering
        \includegraphics[width=\textwidth]{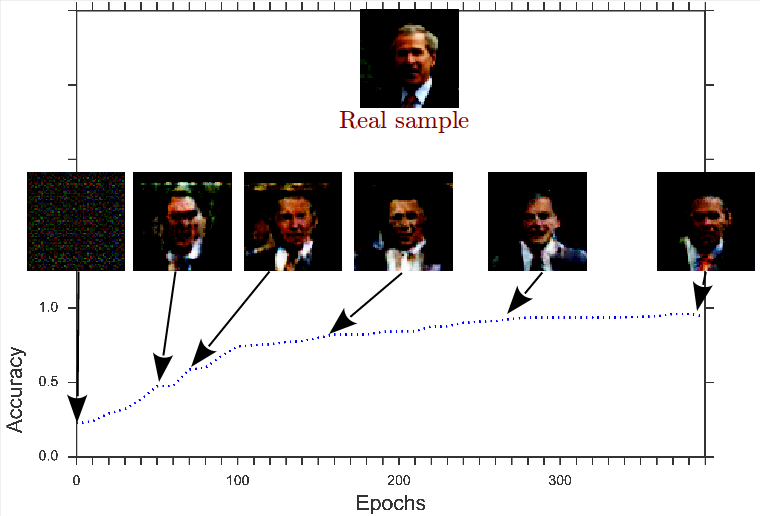}
        \caption{White-box attack}
        \label{fig:faces_wb-lfw-top10}
    \end{subfigure}~~~
   \begin{subfigure}[b]{\subb\textwidth}
        \centering
        \includegraphics[width=\textwidth]{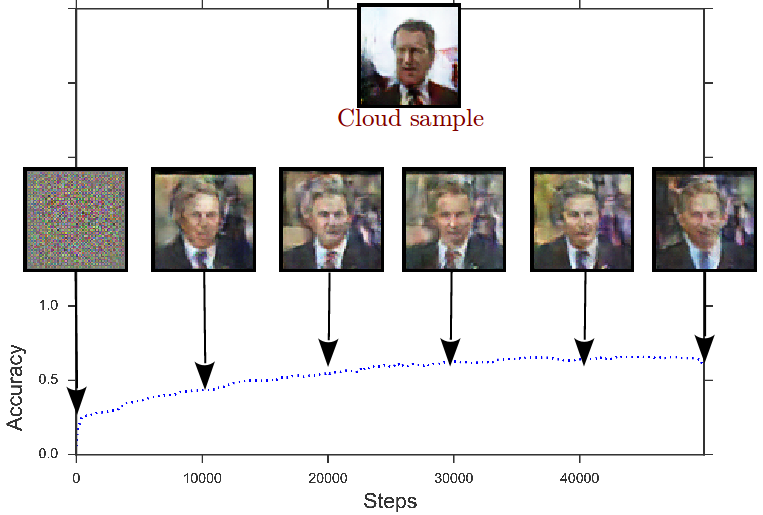}
        \caption{Black-box attack}
        \label{fig:faces_bb-lfw-top10}
    \end{subfigure}
  \vspace{-0.3cm}
    \caption{Accuracy curves and samples at different stages of training on top ten classes from the LFW dataset, showing a clear correlation between higher accuracy and better sample quality.}
    \label{fig:lfw_faces}
\end{figure*}

\descr{Generative approach.} We consider the same set of experiments with similar settings for attacker knowledge as in the discriminative approach; %
the only difference is that in one of the  settings we now assume the attacker has 20\% knowledge of the training set rather than the test set. We
use DCGAN as the generative attacker model. Specifically, we consider that the attacker has:\smallskip

\begin{enumerate}
\item[(1)] 20\% knowledge of the {\em training} set; or
\item[(2)] 30\% knowledge of both the training and test set.\smallskip
\end{enumerate}
In all the experiments, we introduce a delay of 1000 training steps before the attacker model uses
the auxiliary attacker knowledge.
Introducing the auxiliary knowledge early in training process of the attacker model resulted in a weaker discriminator -- see Fig.~\ref{fig:lfw-delay-delta}.

In Fig.~\ref{fig:bb-attack-with-adv-train-gen}, we plot results for setting (1): clearly, there is a substantial increase in accuracy
for the LFW dataset, from 40\% attack accuracy to nearly 60\%. However, there is no increase in accuracy for the CIFAR-10 dataset. Thus, we conclude that setting (1) does
not generalize. Fig.~\ref{fig:bb-attack-with-adv-train-adv-test-gen} shows results for setting (2); for both LFW and CIFAR-10 there is a substantial improvement in accuracy. Accuracy
for the LFW experiment increases from 40\% (with no auxiliary attacker knowledge) to 60\%, while, for CIFAR-10, from 37\% to
58\%. 

Thus, we conclude that, even a small amount of auxiliary attacker knowledge can lead to greatly improving membership inference attacks.

\subsection{Training Performance}\label{sec:remarks}
We also set out to better understand the relationship between membership inference and training performance.
To this end, we report, in Fig.~\ref{fig:lfw_faces}, the attack accuracy and samples generated at different training stages by the target DCGAN generator in the white-box attack (Fig.~\ref{fig:faces_wb-lfw-top10}) 
and the attacker DCGAN generator in the black-box attack (Fig.~\ref{fig:faces_bb-lfw-top10}) on the top ten classes from the LFW dataset. 
The plots demonstrate that accuracy correlates well with the visual quality of the generated samples. 
In particular, samples generated by the target yield a better visual quality than the ones generated by the attacker generator during the black-box attack, and this results in higher membership inference accuracies. 
Overall, the samples generated by both attacks at later stages look visually pleasant, and fairly similar to the original ones.

Our attacks have been evaluated on datasets that consist of complex representations of faces (LFW) and objects (CIFAR-10).
In Appendix~\ref{app:samples}, we include real and generated samples in multiple settings; see Figures~\ref{fig:diab_ret_samples}--\ref{fig:attacker-samples}.
In particular, as shown in Fig.~\ref{fig:real-lfw-top10-samples}, real samples from LFW  contain rich details both in the foreground and background.
We do not observe any large deviations in images within datasets, excluding that the attack performs well due to some training samples being more easily learned by the model, and
so predicting with higher confidence.
Learning the distribution of such images is a challenging task compared to simple datasets such as MNIST, where samples from each class have extremely similar features.
In fact, our black-box attack is able to generate realistic samples (see differences between the target model samples in Fig.~\ref{fig:cloud-dcgan-lfw-top10-samples} and the attacker samples in Fig.~\ref{fig:attacker-dcgan-cloud-dcgan-lfw-top10-samples}).

\begin{figure*}[t]
   \centering
   \begin{subfigure}[b]{.42\textwidth}
        \centering
        \includegraphics[width=\textwidth]{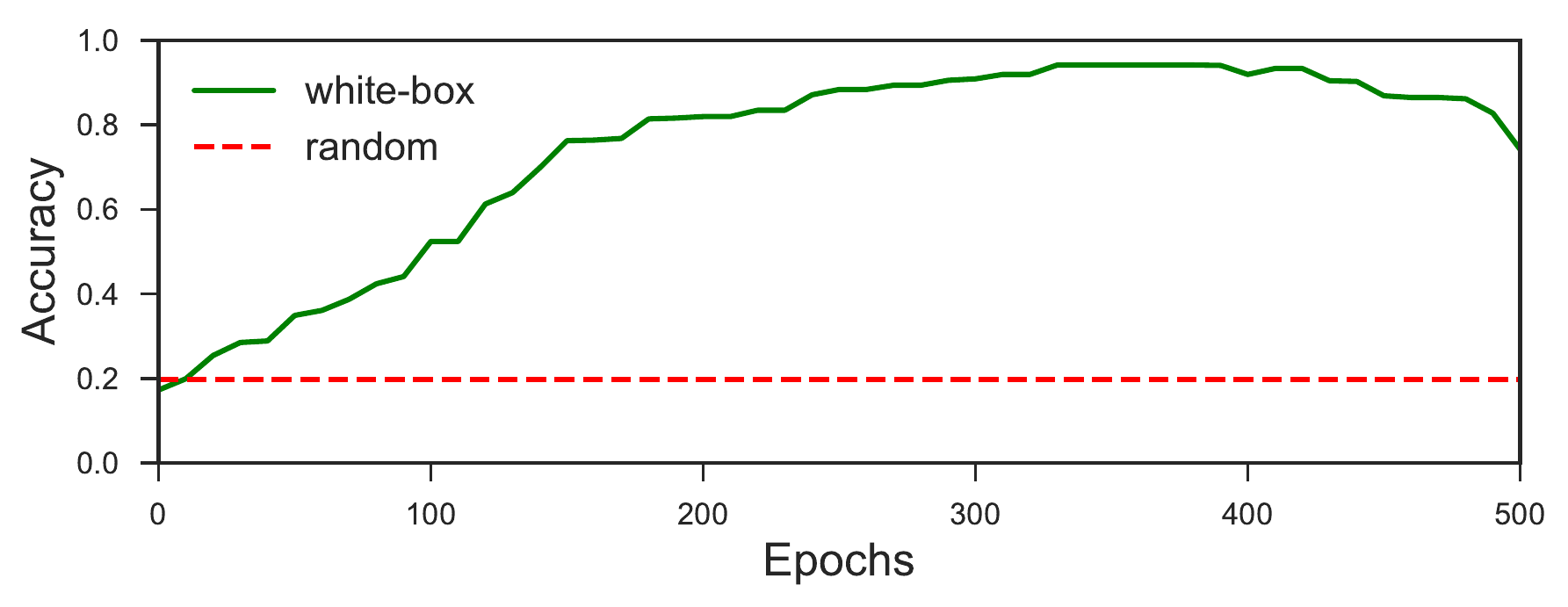}
        \caption{White-box attack}
        \label{fig:acc_wb_diab}
    \end{subfigure}
   \begin{subfigure}[b]{.42\textwidth}
        \centering
        \includegraphics[width=\textwidth]{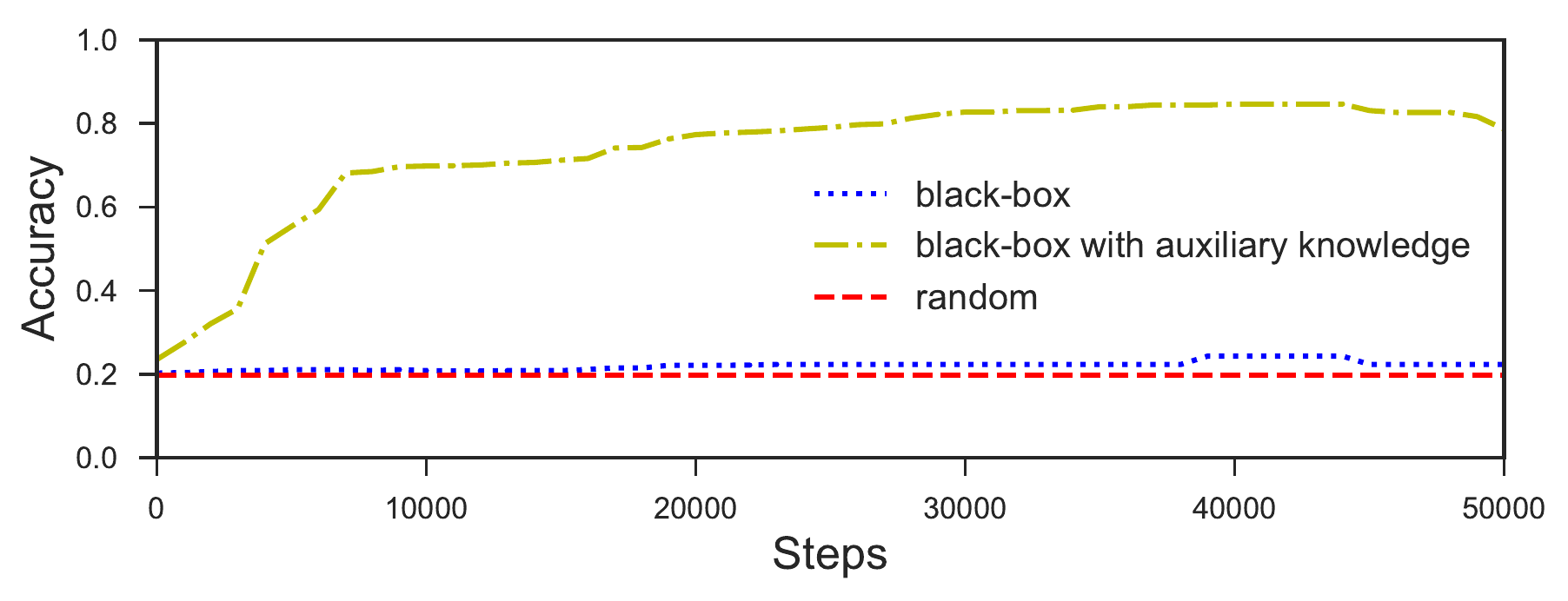}
        \caption{Black-box attack}
        \label{fig:acc_bb_diab}
    \end{subfigure}
\vspace{-0.2cm}
    \caption{Accuracy curves of attacks against a DCGAN target model on the Diabetic Retinopathy dataset.}
    \label{fig:diab_ret}
   \vspace{-0.2cm}    
\end{figure*}

\subsection{Evaluation on Diabetic Retinopathy Dataset}\label{sec:eval_eye}

Finally, we present a case study of our attacks on the Diabetic Retinopathy (DR) dataset, which consists of high-resolution retina images, 
with an integer label assigning a score of the degree to which the participant suffers from diabetic retinopathy.
Diabetic  retinopathy is a leading cause of blindness in the developed world, with detection currently performed manually by highly skilled clinicians. The machine learning competition site \url{kaggle.com} has evaluated proposals for automated detection of diabetic retinopathy, and submissions have demonstrated high accuracies.
of manual detection.  %

We choose this additional dataset since the generation of synthetic medical images through generative models is a powerful method to produce large numbers of high-quality sample data on which useful machine learning models can be trained. Thus, our attacks raise serious privacy concerns, in practice, in such sensitive settings as they involve (sensitive) medical data.

As discussed in Section~\ref{ssec:experimental}, the dataset includes 88,702 high-resolution retina images under various imaging conditions. 
Each image is labelled with an integer representing how present is diabetic retinopathy within the retina, from 0 to 4. We train the generative target model on images with labels 2, 3 and 4, i.e., with mild to severe cases of diabetic retinopathy. These make up 19.7\% of the dataset.
(Fig.~\ref{fig:diab_ret_samples} in Appendix~\ref{app:samples} show real and target generated samples of retina images.)

The results of the white-box attack are reported in Fig.~\ref{fig:acc_wb_diab}: the attack is overwhelmingly successful, nearing 100\% accuracy at 350 training epochs. 
Fig.~\ref{fig:acc_bb_diab} shows the black-box attacks results, when an attacker
has no auxiliary knowledge, and when the attacker has 30\% training and test set auxiliary knowledge. A no-knowledge black-box attack does not perform very well, while, with some auxiliary knowledge, it approaches the accuracy of the white-box attack, peaking at over 80\% after 35K training steps.

\section{Discussion}\label{sec:discussion}

In this section, we summarize our results, then, measure the sensitivity of the attacks to training set size and prediction ordering.
Finally, we study robustness to possible defenses.

\begin{figure*}[t]
   \centering
   \begin{subfigure}[b]{\sube\textwidth}
        \centering
        \includegraphics[width=\textwidth]{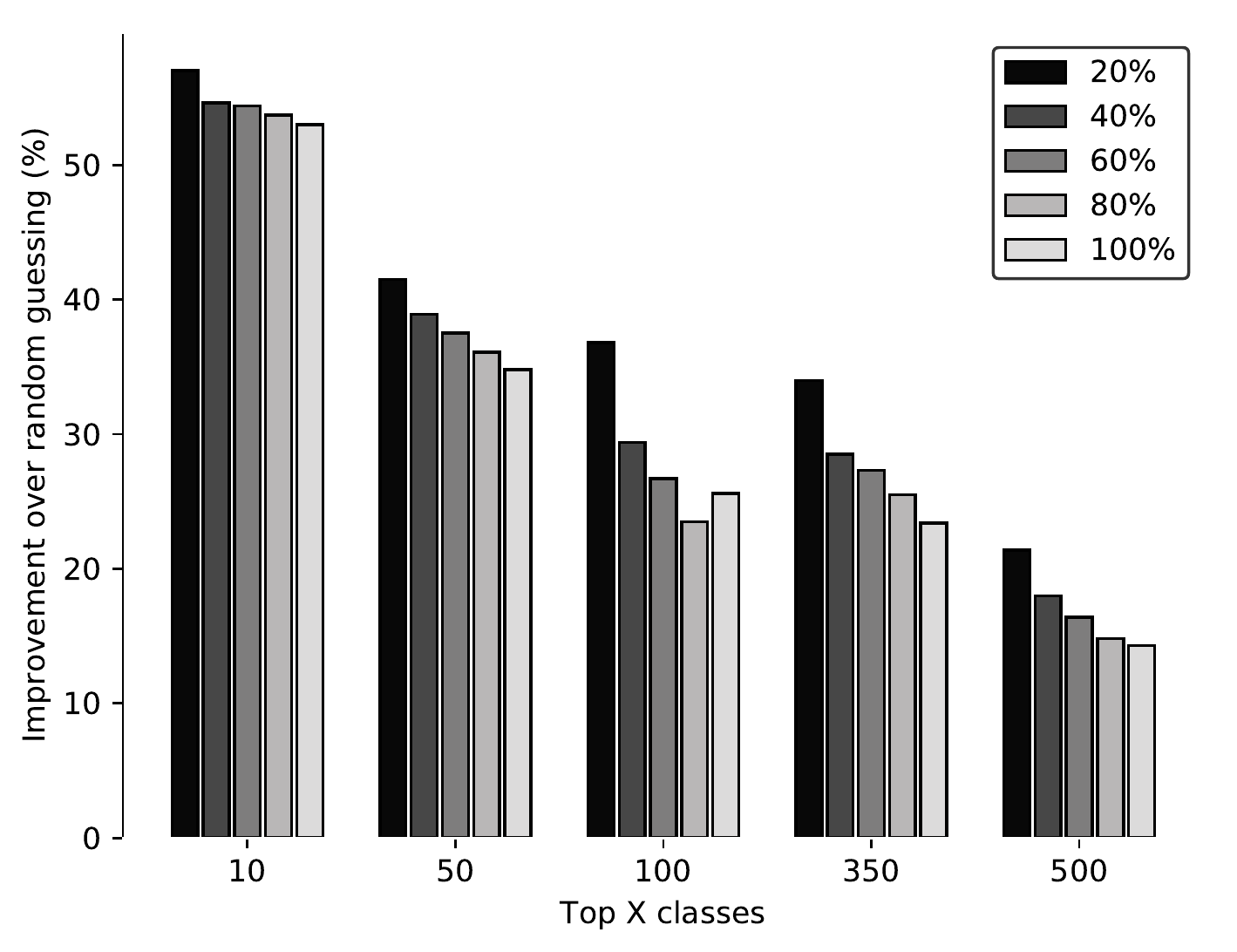}
        \caption{LFW Top X classes}
        \label{fig:bb_lfw_top_size_exp}
    \end{subfigure}
    ~~
    \begin{subfigure}[b]{\sube\textwidth}
        \centering
        \includegraphics[width=\textwidth]{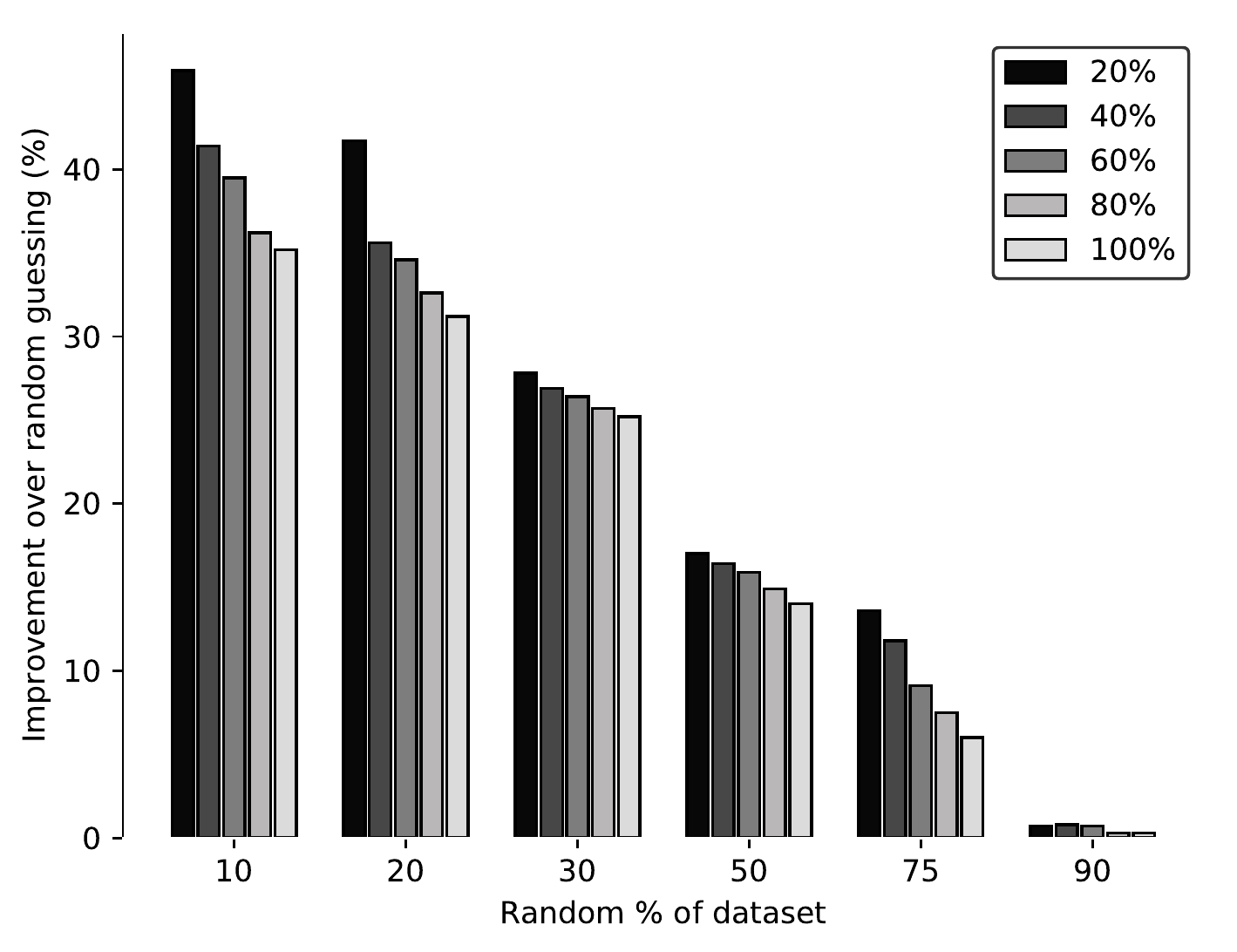}
        \caption{LFW, random X\% subset}
        \label{fig:bb_lfw_rand_size_exp}
    \end{subfigure}
    ~~
    \begin{subfigure}[b]{\sube\textwidth}
        \centering
        \includegraphics[width=\textwidth]{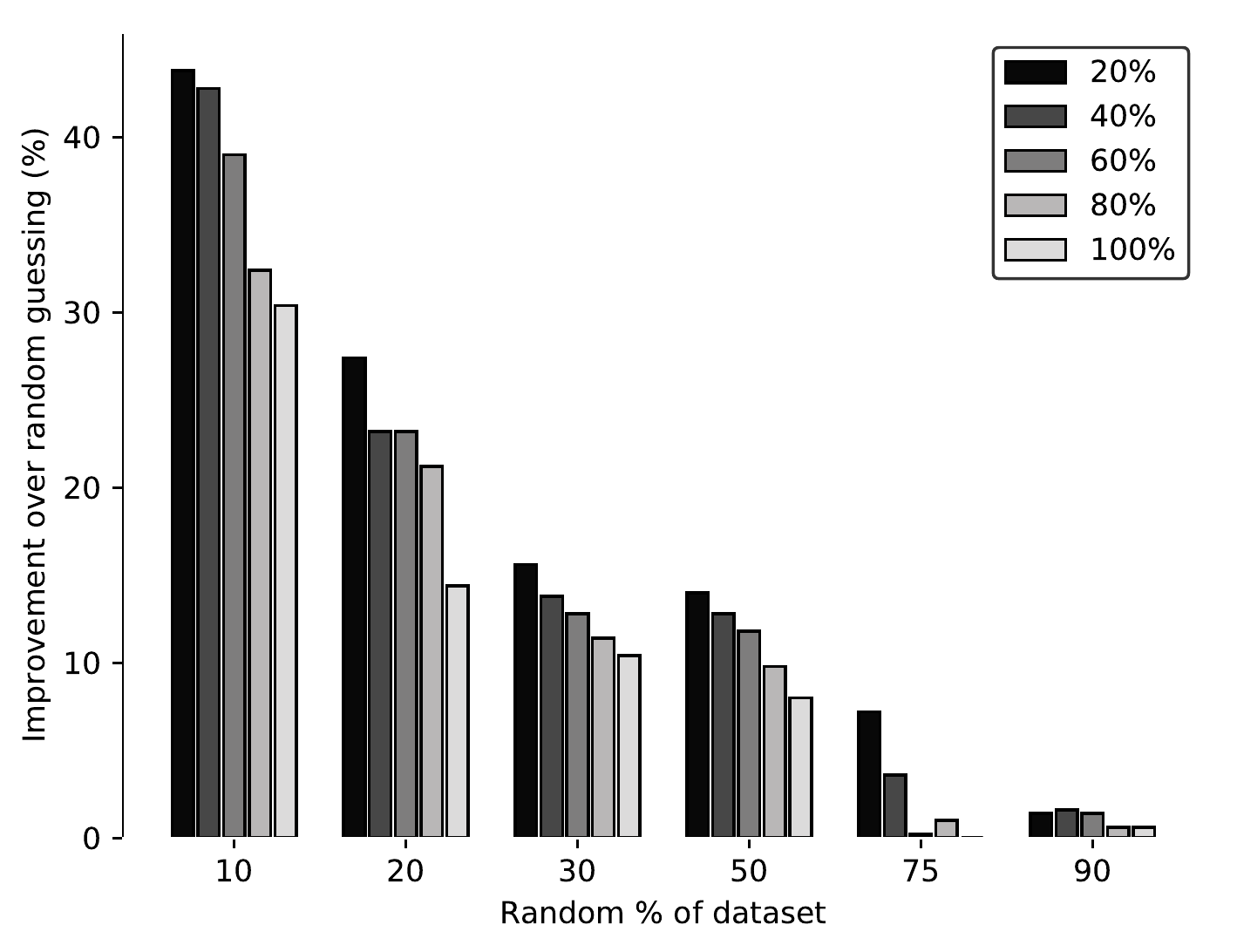}
        \caption{CIFAR-10 random X\% subset}
        \label{fig:bb_cifar10_rand_size_exp}
    \end{subfigure}
    \vspace{-0.2cm}
    \caption{Improvements over random guessing, in a black-box attack, as we vary the size of the training set, and consider smaller subsets for training set predictions.}
    \label{fig:size_exp}
    \vspace{-0.2cm}
\end{figure*}

\subsection{Summary of Results}

Overall, our analysis shows that state-of-the-art generative models are vulnerable against membership inference attacks. 
In Table~\ref{tab:attack_results_summary}, we summarize the best accuracy results for experiments on random 10\% training sets (LFW, CIFAR-10) and the diabetic retinopathy (DR) dataset experiments.

We note that, for white-box attacks, the attacker successfully infers the training set with 100\% accuracy on both the LFW and CIFAR-10 datasets, and 95\% accuracy for DR dataset.
Accuracy drops to 40\% on LFW, 37\% on CIFAR-10 and 22\% on DR for black-box attacks with no auxiliary knowledge, 
however, even with a small amount of auxiliary knowledge, the attacker boost performance up to 60\% on LFW, 58\% on CIFAR-10 and 81\% on DR. 
Note that a random guess corresponds to 10\% accuracy on LFW and CIFAR-10, and 20\% on DR.
Further, we show that our attacks are robust against different target model architectures.

\begin{table}[t]
   \begin{center}
   \setlength{\tabcolsep}{4pt}
        \begin{small}
                    \begin{tabular}{lrrr}
                    \toprule
                        {\bf Attack}      &  {\bf   LFW}         &         {\bf CIFAR-10} & {\bf DR}              \\
						\midrule
                        White-box                              & 100\% & 100\% & 95\% \\ 
                        Black-box with no knowledge         & 40\% & 37\%  & 22\% \\
                        Black-box with limited knowledge       & 60\% & 58\%  & 81\%  \\
                        {\em Random Guess}						& 10\% & 10\% & 20\%\\
                        \bottomrule
                  \end{tabular}
        \end{small}
   \end{center}
   \vspace{-0.4cm}
   \caption{Accuracy of the best attacks on random 10\% training set for LFW and CIFAR-10, and for diabetic retinopathy (DR).}
    \label{tab:attack_results_summary}
   \vspace{-0.5cm}
\end{table}

\subsection{Sensitivity to training set size and prediction ordering}\label{sec:attack_sizes}

Aiming to measure the dependency between attack performance and training set size, we experiment
with varying training set sizes in the DCGAN target and attacker model setting.

Fig.~\ref{fig:size_exp} shows how the improvement of the attack
degrades as the relative size of the training set increases. Note that we only include black-box attack results, as \textit{all white-box attacks achieve
almost 100\% accuracy regardless of training set size.}
Overall, we find that there is a commonality in the experiments:
black-box attacks on 10\% of the dataset achieve an improvement of 40--55\%, and, as we increase the number of data-points used to train the target model, the attack has smaller and smaller improvements over random guessing.

The largest increases are in the setting of Fig.~\ref{fig:bb_lfw_top_size_exp},
where data-points are more homogeneous and so overfitting effects are compounded. When the training set is 90\% of the total dataset used in the
evaluation of the attack, the attack has negligible improvements over random guessing. 
We believe that this might be due either to:
(1) the larger number of training data-points yields a well-fitted model that does not leak information about training records, or (2) a small number of data-points within
the training set do not leak information, therefore, as we increase the size of the training set, the inability to capture these records becomes
more costly, resulting in smaller improvements in attack performance.

If the former were true, we would see smaller improvements for larger training sets, regardless of the total size of the dataset; however, experiments on both LFW and CIFAR-10, which consist of different training sizes, report similar improvements over random guessing. Additionally, white-box
attacks are not affected by increasing the training set size, which would be the case if the model did not overfit and thus leak information about training records. 
Hence, we believe a small number of training records are inherently difficult to capture, and so  improvements over random guessing for larger training set sizes are more difficult to achieve since the majority of samples are used to train the target model.

We also examine the attack sensitivity to the ordering of the data-point predictions. So far, the \emph{only} prior knowledge
the attacker has is the approximate size of the training set. If there is a clear ordering of data-points predictions, with training records sitting
at the top of the ordering, and non-training records lower down, an attacker can use this information to identify training records without side knowledge of training set size.
They can simply place a confidence score relative to where in the ordering a data-point predictions sits.

 Fig.~\ref{fig:size_exp} shows, for varying training set sizes,
how many training records lie in the top 20\%, 40\%, 60\%, 80\%, and 100\% of the guessed training set. We observe that, in all experimental settings,
accuracy for the top 20\% is highest, with scores decreasing as the attacker considers a larger number of data-points as candidates for
the training set. 

Thus, training to non-training samples follows a structured ordering in the attacker's predictions, which can be exploited to infer membership when the attacker has \emph{no knowledge} of the original training set size by setting a threshold on the minimum confidence of a training point.

\subsection{Defenses} %
\label{ssec:defenses}

Possible defense strategies against membership inference (see~\cite{shokri2016membership}), e.g., restricting the prediction vector to the top $k$ classes, coarsening and increasing the entropy of the prediction vector, are not well suited to our attacks, since generative models do not output prediction vectors.
However, regularization techniques and differential privacy could possibly be applied to generative models to produce more robust and stable training as well as more diverse and visually pleasant samples. %

\descr{Weight Normalization and Dropout.} To this end, we consider two techniques, namely, Weight Normalization~\cite{salimans2016weight} and 
Dropout~\cite{srivastava2014dropout}, as possible defense mechanisms and evaluate their impact on our 
attacks.\footnote{Note that we do not compare models with and without Batch Normalization~\cite{ioffe2015batch}, as its inclusion has shown to improve sample quality and is nearly always used in model construction of GANs~\cite{radford2015unsupervised}.}
The former is a re-parameterization of the weights vectors that decouples the length of those weights from their direction, and it is applied to all layers in both generator and discriminator in the target model.
Whereas, the latter can be used to prevent overfitting by randomly dropping out (i.e., zeroing) connections between neurons during training---in particular, we apply Dropout, with probability 0.5, to all the layers in the discriminator.

\begin{figure}[t]
   \centering
   \includegraphics[width=0.36\textwidth]{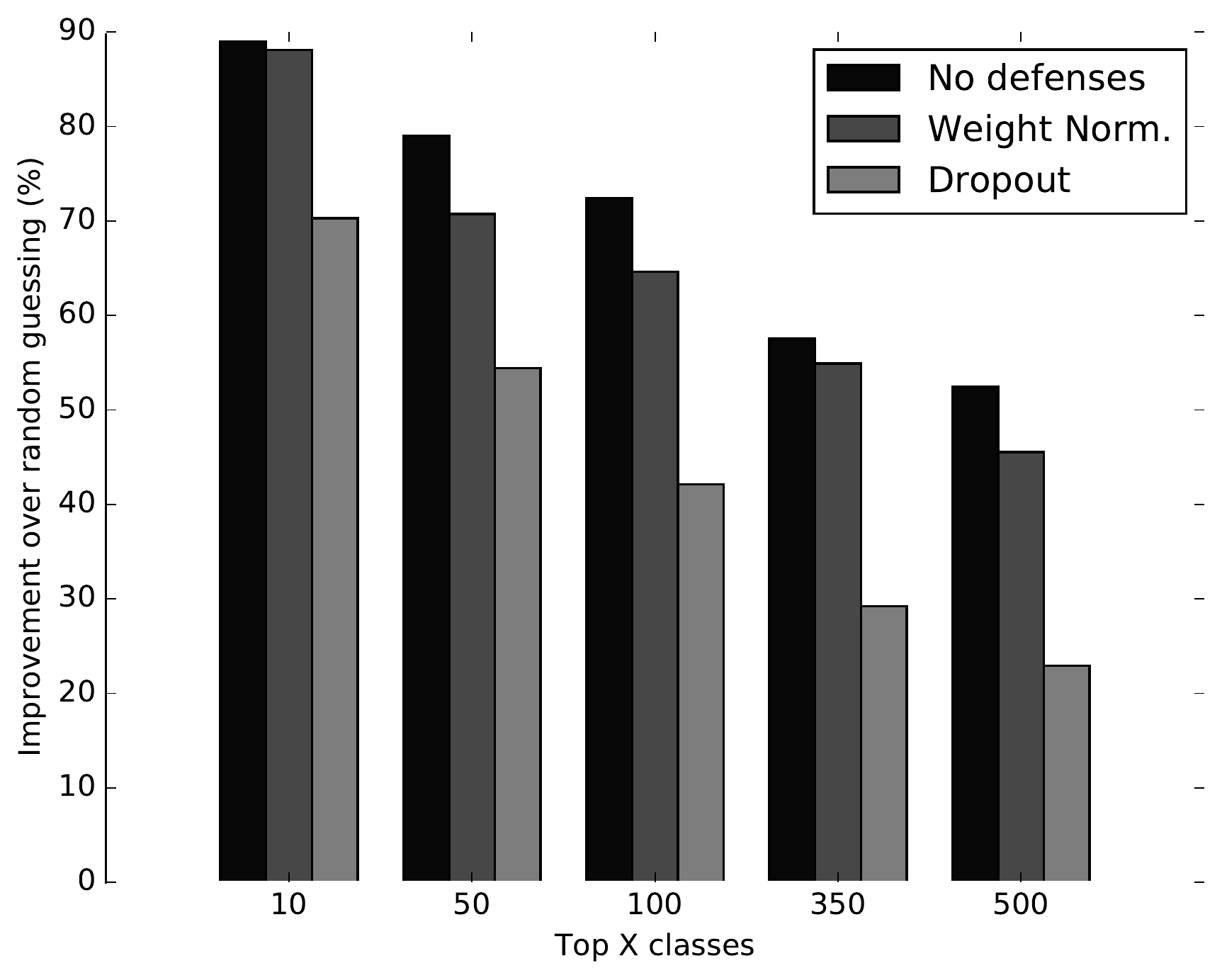}
   \vspace{-0.2cm}   
   \caption{{Improvement over random guessing for Weight Normalization and Dropout defenses against white-box attacks on models trained over different number of classes with LFW.}}
   \label{fig:lfw-defenses}
   \vspace{-0.2cm}   
\end{figure}

In Fig.~\ref{fig:lfw-defenses},  we measure the improvement over random guessing for the white-box attack against the target model trained on LFW using either Weight Normalization or Dropout.
With Weight Normalization, we get improvements over random guessing of, respectively, 88\% and 46\%, which are very close to the target model trained with no defenses (resp., 89\% and 52\%).  
Dropout is more effective, as the improvements over random guessing go down to 70\% on top 10 classes and 23\% on top 500 classes.

However, Dropout significantly slows down the training process, requiring more epochs to get qualitatively plausible samples.
Also, Weight Normalization often results in training instability (i.e., the discriminator outperforms the generator, or vice-versa).

\begin{figure}[t]
   \centering
   \includegraphics[width=0.438\textwidth]{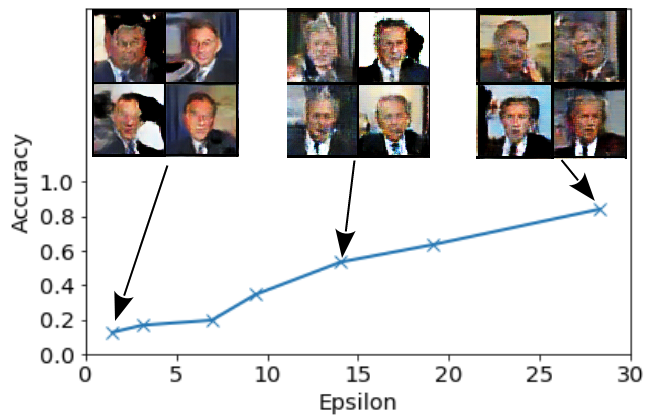}
   \caption{{Accuracy curve and samples for different privacy budgets on top ten classes from the LFW dataset, showing a trade-off between samples quality and privacy guarantees.}}
   \label{fig:dp-defenses}
   \vspace{-0.2cm}
\end{figure}

\descr{Differentially Private GANs.} 
We also evaluate our attack against a recently proposed technique for $(\varepsilon,\delta)$-Differentially Private GANs~\cite{triastcyn2018generating}, where Gaussian noise~\cite{dwork2008differential} is injected in the discriminator forward pass during training.
Fig.~\ref{fig:dp-defenses} shows the results of a white-box attack against Differentially Private DCGAN trained on top ten classes for different values of the privacy budget $\varepsilon$ (with $\delta$ set to $10^{-4}$). 
For all experiments, the target model is trained for $500$ epochs and the final privacy budget is computed using moments accountant~\cite{abadi2016deep}.  
The attack does no better than random guessing for $\varepsilon=1.5$ (first tick in the plot), while accuracy increases up to 85\% for $\varepsilon=28.3$. 
However, note that acceptable levels of privacy (i.e., values of $\varepsilon < 10$) yield very bad samples quality.

\descr{Using our attacks as defense.} Also note that, as discussed in Section~\ref{sec:intro}, our attacks can actually be used as a defense mechanism. 
The difference in white-box and black-box accuracy provides information
about how well the local model approximates the target model, thus, one could use this information to train a target model which cannot be well approximated. Furthermore, similarly to {\em early-stopping} criteria in model training, one can stop training when visual sample quality is high but white-box attack accuracy is still low.

In our experiments, we also observe the benefits of a more regularized model in increasing the robustness against information leakage in the case of BEGAN.
For instance, in white-box attacks on CIFAR-10, BEGAN produces quality samples without overfitting, with membership inference performing only 9\% better than random guessing (see~Fig.~\ref{fig:smoothed-wb-cifar-rand10}).

\subsection{Cost of the Attacks}\label{app:cost}

Finally, we quantify the cost of the attacks in terms of computational and time overhead, and estimate monetary costs.

To perform the attacks, the attacker needs a GPU, which can be obtained for a cost in the order of \$100.
The attacks have minimal running time overheads: for the white-box attack, complexity is negligible as we only query a pre-trained target model to steal discriminator model parameters, whereas, for black-box, one step of training the attacker model takes 0.05 seconds in our testbed. 
Black-box attacks with no auxiliary attacker knowledge yield the best results after 50,000 training steps, therefore, an attacker can expect best results after approximately 42 minutes 
with 32 $\times$ 50,000 queries to the target model (since we define one training step as one mini-batch iteration, with 32 inputs per mini-batch). 
For attacks with auxiliary knowledge, the best results are reached after 15,000 training steps, thus, approximately 13 minutes. %

We also estimate monetary cost based on current discriminative MLaaS pricing structures from Google.\footnote{\url{https://cloud.google.com/vision/pricing}} At a cost of \$1.50 per 1,000 target queries, after an initial 1,000 free monthly queries, the black-box attack with no auxiliary knowledge would cost \$2,352, while the black-box attack with auxiliary knowledge \$672.
Therefore, we consider our attacks to have minimal costs, especially considering the potential severity of the information leakage they enable.

\section{Conclusion}\label{sec:conclusion}

This paper presented the first evaluation of membership inference attacks against generative models, showing that a variety of models lead to important privacy leakage.
Our attacks are cheap to run, do not need information about the model under attack, and generalize well.
Moreover, membership inference is harder to mount on generative models than it is on discriminative ones; in the latter, the attacker can use the confidence the model places on an input belonging to a label to perform the attack, while in the former there is no such signal.

We conducted an experimental evaluation on state-of-the-art probabilistic models such as Deep Convolutional GAN (DCGAN), Boundary Equilibrium GAN (BEGAN), and the combination of DCGAN with a Variational Autoencoder (DCGAN+VAE), using datasets with complex representations of faces (LFW), objects (CIFAR-10), and medical images with real-world privacy concerns (Diabetic Retinopathy).
We showed that the white-box attack can be used to detect overfitting in generative models and help selecting an appropriate model that will not leak information about samples on which it was trained.
We also demonstrated that our low-cost black-box attack can perform membership inference using a novel method for
training GANs, and that an attacker with limited auxiliary knowledge of dataset samples can remarkably improve their accuracy. %

Moreover, we experimented with regularization techniques, such as Weight Normalization~\cite{salimans2016weight} and Dropout~\cite{srivastava2014dropout}, and differentially private mechanisms, which could be used to mitigate our attacks. 
We found that they are effective up to a certain extent, but need longer training, yield training instability, and/or worse generated samples (in terms of quality).
This motivates the need for future work on defenses against information leakage in generative models.

Our work also provides evidence that models that generalize well (e.g., BEGAN) yield higher protection against membership inference attacks, confirming that generalization and privacy are associated. 
Thus, our evaluation may be used to empirically assess the generalization quality of a generative model, which is an open research problem of independent interest.
As part of future work, we plan to apply our attacks to other privacy-sensitive datasets, including location data. 

\descr{Acknowledgments.} This work was partially supported by The Alan Turing Institute under the EPSRC grant EP/N510129/1 and a grant by Nokia Bell Labs. Jamie Hayes is supported by a Google PhD Fellowship in Machine Learning.

\bibliographystyle{abbrv}

\begin{thebibliography}{10}

\bibitem{abadi2016deep}
M.~Abadi, A.~Chu, I.~Goodfellow, H.~B. McMahan, I.~Mironov, K.~Talwar, and
  L.~Zhang.
\newblock {Deep learning with differential privacy}.
\newblock In {\em CCS}, 2016.

\bibitem{aono2017privacy}
Y.~Aono, T.~Hayashi, L.~Wang, S.~Moriai, et~al.
\newblock {Privacy-preserving deep learning: Revisited and Enhanced}.
\newblock In {\em ATIS}, 2017.

\bibitem{arjovsky2017wasserstein}
M.~Arjovsky, S.~Chintala, and L.~Bottou.
\newblock {Wasserstein GAN}.
\newblock {\em arXiv 1701.07875}, 2017.

\bibitem{ateniese2015hacking}
G.~Ateniese, L.~V. Mancini, A.~Spognardi, A.~Villani, D.~Vitali, and G.~Felici.
\newblock {Hacking smart machines with smarter ones: How to extract meaningful
  data from machine learning classifiers}.
\newblock {\em International Journal of Security and Networks}, 2015.

\bibitem{backes2016membership}
M.~Backes, P.~Berrang, M.~Humbert, and P.~Manoharan.
\newblock {Membership Privacy in MicroRNA-based Studies}.
\newblock In {\em CCS}, 2016.

\bibitem{beaulieu2017privacy}
B.~K. Beaulieu-Jones, Z.~S. Wu, C.~Williams, and C.~S. Greene.
\newblock Privacy-preserving generative deep neural networks support clinical
  data sharing.
\newblock {\em bioRxiv}, 2017.

\bibitem{bengio2013generalized}
Y.~Bengio, L.~Yao, G.~Alain, and P.~Vincent.
\newblock {Generalized denoising auto-encoders as generative models}.
\newblock In {\em NIPS}, 2013.

\bibitem{berthelot2017began}
D.~Berthelot, T.~Schumm, and L.~Metz.
\newblock {BEGAN: Boundary Equilibrium Generative Adversarial Networks}.
\newblock {\em arXiv 1703.10717}, 2017.

\bibitem{bonawitzpractical}
K.~Bonawitz, V.~Ivanov, B.~Kreuter, A.~Marcedone, H.~B. McMahan, S.~Patel,
  D.~Ramage, A.~Segal, and K.~Seth.
\newblock Practical secure aggregation for privacy preserving machine learning.
\newblock In {\em CCS}, 2017.

\bibitem{calandrino2011you}
J.~A. Calandrino, A.~Kilzer, A.~Narayanan, E.~W. Felten, and V.~Shmatikov.
\newblock {``You Might Also Like:'' Privacy Risks of Collaborative Filtering}.
\newblock In {\em IEEE Security and Privacy}, 2011.

\bibitem{carlini2018secret}
N.~Carlini, C.~Liu, J.~Kos, {\'U}.~Erlingsson, and D.~Song.
\newblock {The Secret Sharer: Measuring Unintended Neural Network Memorization
  \& Extracting Secrets}.
\newblock {\em arXiv:1802.08232}, 2018.

\bibitem{ganhacks}
S.~Chintala, E.~Denton, M.~Arjovsky, and M.~Mathieu.
\newblock {How to Train a GAN? Tips and tricks to make GANs work}.
\newblock \url{https://github.com/soumith/ganhacks}, Year.

\bibitem{choi17generating}
E.~{Choi}, S.~{Biswal}, B.~{Malin}, J.~{Duke}, W.~F. {Stewart}, and J.~{Sun}.
\newblock {Generating Multi-label Discrete Electronic Health Records using
  Generative Adversarial Networks}.
\newblock In {\em Machine Learning for Healthcare}, 2017.

\bibitem{dowlin2016cryptonets}
N.~Dowlin, R.~Gilad-Bachrach, K.~Laine, K.~Lauter, M.~Naehrig, and J.~Wernsing.
\newblock {Cryptonets: Applying neural networks to encrypted data with high
  throughput and accuracy}.
\newblock In {\em ICML}, 2016.

\bibitem{du2004privacy}
W.~Du, Y.~S. Han, and S.~Chen.
\newblock {Privacy-preserving multivariate statistical analysis: Linear
  regression and classification}.
\newblock In {\em ICDM}, 2004.

\bibitem{dwork2008differential}
C.~Dwork.
\newblock {Differential privacy: A survey of results}.
\newblock In {\em Theory and Applications of Models of Computation}, 2008.

\bibitem{dwork2015generalization}
C.~Dwork, V.~Feldman, M.~Hardt, T.~Pitassi, O.~Reingold, and A.~Roth.
\newblock Generalization in adaptive data analysis and holdout reuse.
\newblock In {\em NIPS}, 2015.

\bibitem{fredrikson2015model}
M.~Fredrikson, S.~Jha, and T.~Ristenpart.
\newblock {Model inversion attacks that exploit confidence information and
  basic countermeasures}.
\newblock In {\em CCS}, 2015.

\bibitem{fredrikson2014privacy}
M.~Fredrikson, E.~Lantz, S.~Jha, S.~Lin, D.~Page, and T.~Ristenpart.
\newblock {Privacy in pharmacogenetics: An end-to-end case study of
  personalized warfarin dosing}.
\newblock In {\em USENIX Security}, 2014.

\bibitem{goodfellow2014generative}
I.~Goodfellow, J.~Pouget-Abadie, M.~Mirza, B.~Xu, D.~Warde-Farley, S.~Ozair,
  A.~Courville, and Y.~Bengio.
\newblock {Generative adversarial nets}.
\newblock In {\em NIPS}, 2014.

\bibitem{gulrajani2017improved}
I.~Gulrajani, F.~Ahmed, M.~Arjovsky, V.~Dumoulin, and A.~Courville.
\newblock {Improved training of Wasserstein GANs}.
\newblock In {\em ICLR (Posters)}, 2018.

\bibitem{hinton2015distilling}
G.~Hinton, O.~Vinyals, and J.~Dean.
\newblock Distilling the knowledge in a neural network.
\newblock {\em arXiv 1503.02531}, 2015.

\bibitem{hitaj2017deep}
B.~Hitaj, G.~Ateniese, and F.~Perez-Cruz.
\newblock {Deep Models Under the GAN: Information Leakage from Collaborative
  Deep Learning}.
\newblock In {\em CCS}, 2017.

\bibitem{homer2008resolving}
N.~Homer, S.~Szelinger, M.~Redman, D.~Duggan, W.~Tembe, J.~Muehling, J.~V.
  Pearson, D.~A. Stephan, S.~F. Nelson, and D.~W. Craig.
\newblock {Resolving individuals contributing trace amounts of DNA to highly
  complex mixtures using high-density SNP genotyping microarrays}.
\newblock {\em PLoS Genet}, 2008.

\bibitem{LFWTech}
G.~B. Huang, M.~Ramesh, T.~Berg, and E.~Learned-Miller.
\newblock {Labeled Faces in the Wild: A Database for Studying Face Recognition
  in Unconstrained Environments}.
\newblock Technical report, University of Massachusetts, Amherst, 2007.
\newblock \url{http://vis-www.cs.umass.edu/lfw/lfw.pdf}.

\bibitem{ioffe2015batch}
S.~Ioffe and C.~Szegedy.
\newblock Batch normalization: Accelerating deep network training by reducing
  internal covariate shift.
\newblock In {\em International Conference on Machine Learning}, 2015.

\bibitem{ji2015your}
S.~Ji, W.~Li, N.~Z. Gong, P.~Mittal, and R.~A. Beyah.
\newblock On your social network de-anonymizablity: Quantification and large
  scale evaluation with seed knowledge.
\newblock In {\em NDSS}, 2015.

\bibitem{jia2018attriguard}
J.~Jia and N.~Z. Gong.
\newblock Attriguard: A practical defense against attribute inference attacks
  via adversarial machine learning.
\newblock In {\em USENIX Security}, 2018.

\bibitem{dr}
{Kaggle.com}.
\newblock {Diabetic Retinopathy Detection}.
\newblock
  \url{https://www.kaggle.com/c/diabetic-retinopathy-detection#references},
  2015.

\bibitem{generative}
A.~Karpathy, P.~Abbeel, G.~Brockman, P.~Chen, V.~Cheung, R.~Duan,
  I.~Goodfellow, D.~Kingma, J.~Ho, R.~Houthooft, T.~Salimans, J.~Schulman,
  I.~Sutskever, and W.~Zaremba.
\newblock {Generative Models}.
\newblock \url{https://blog.openai.com/generative-models/}, 2017.

\bibitem{kingma2013auto}
D.~P. Kingma and M.~Welling.
\newblock {Auto-Encoding Variational Bayes}.
\newblock In {\em ICLR}, 2013.

\bibitem{krizhevsky2009learning}
A.~Krizhevsky and G.~Hinton.
\newblock Learning multiple layers of features from tiny images.
\newblock Technical report, University of Toronto, 2009.
\newblock \url{https://www.cs.toronto.edu/~kriz/learning-features-2009-TR.pdf}.

\bibitem{kusner2015differentially}
M.~J. Kusner, J.~R. Gardner, R.~Garnett, and K.~Q. Weinberger.
\newblock {Differentially Private Bayesian Optimization}.
\newblock In {\em ICML}, 2015.

\bibitem{larsen2015autoencoding}
A.~B.~L. Larsen, S.~K. S{\o}nderby, H.~Larochelle, and O.~Winther.
\newblock {Autoencoding beyond pixels using a learned similarity metric}.
\newblock In {\em ICLM}, 2016.

\bibitem{ledig2016photo}
C.~Ledig, L.~Theis, F.~Husz{\'a}r, J.~Caballero, A.~Cunningham, A.~Acosta,
  A.~Aitken, A.~Tejani, J.~Totz, Z.~Wang, et~al.
\newblock {Photo-realistic single image super-resolution using a generative
  adversarial network}.
\newblock arXiv 1609.04802, 2016.

\bibitem{lindell2000privacy}
Y.~Lindell and B.~Pinkas.
\newblock {Privacy preserving data mining}.
\newblock In {\em {CRYPTO}}, 2000.

\bibitem{long2018understanding}
Y.~Long, V.~Bindschaedler, L.~Wang, D.~Bu, X.~Wang, H.~Tang, C.~A. Gunter, and
  K.~Chen.
\newblock {Understanding Membership Inferences on Well-Generalized Learning
  Models}.
\newblock {\em arXiv:1802.04889}, 2018.

\bibitem{lucic2017gans}
M.~{Lucic}, K.~{Kurach}, M.~{Michalski}, S.~{Gelly}, and O.~{Bousquet}.
\newblock {Are GANs Created Equal? A Large-Scale Study}.
\newblock {\em ArXiv 1711.10337}, 2017.

\bibitem{mcmahan2016communication}
H.~B. McMahan, E.~Moore, D.~Ramage, S.~Hampson, et~al.
\newblock Communication-efficient learning of deep networks from decentralized
  data.
\newblock In {\em AISTATS}, 2017.

\bibitem{blogInversion}
F.~McSherry.
\newblock {Statistical inference considered harmful}.
\newblock
  \url{https://github.com/frankmcsherry/blog/blob/master/posts/2016-06-14.md},
  2016.

\bibitem{melis2018inference}
L.~Melis, C.~Song, E.~De~Cristofaro, and V.~Shmatikov.
\newblock {Inference Attacks Against Collaborative Learning}.
\newblock {\em arXiv:1805.04049}, 2018.

\bibitem{narayanan2009anonymizing}
A.~Narayanan and V.~Shmatikov.
\newblock De-anonymizing social networks.
\newblock In {\em IEEE Security and Privacy}, 2009.

\bibitem{nasr2018machine}
M.~Nasr, R.~Shokri, and A.~Houmansadr.
\newblock {Machine Learning with Membership Privacy using Adversarial
  Regularization}.
\newblock In {\em ACM CCS}, 2018.

\bibitem{nie2016medical}
D.~Nie, R.~Trullo, C.~Petitjean, S.~Ruan, and D.~Shen.
\newblock {Medical Image Synthesis with Context-Aware Generative Adversarial
  Networks}.
\newblock In {\em MICCAI}, 2017.

\bibitem{otoro}
{otoro.net}.
\newblock {Generating Large Images from Latent Vectors}.
\newblock
  \url{http://blog.otoro.net/2016/04/01/generating-large-images-from-latent-vectors/},
  2016.

\bibitem{papernot2016semi}
N.~Papernot, M.~Abadi, {\'U}.~Erlingsson, I.~Goodfellow, and K.~Talwar.
\newblock {Semi-supervised knowledge transfer for deep learning from private
  training data}.
\newblock In {\em ICLR}, 2017.

\bibitem{papernot2016distillation}
N.~Papernot, P.~McDaniel, X.~Wu, S.~Jha, and A.~Swami.
\newblock {Distillation as a defense to adversarial perturbations against deep
  neural networks}.
\newblock In {\em IEEE Security and Privacy}, 2016.

\bibitem{papernot2018scalable}
N.~Papernot, S.~Song, I.~Mironov, A.~Raghunathan, K.~Talwar, and
  {\'U}.~Erlingsson.
\newblock {Scalable Private Learning with PATE}.
\newblock In {\em ICLR}, 2018.

\bibitem{pyrgelis2017does}
A.~Pyrgelis, C.~Troncoso, and E.~De~Cristofaro.
\newblock {What Does The Crowd Say About You? Evaluating Aggregation-based
  Location Privacy}.
\newblock In {\em {PETS}}, 2017.

\bibitem{pyrgelis2017knock}
A.~Pyrgelis, C.~Troncoso, and E.~De~Cristofaro.
\newblock {Knock Knock, Who's There? Membership Inference on Aggregate Location
  Data}.
\newblock In {\em NDSS}, 2018.

\bibitem{qian2016anonymizing}
J.~Qian, X.-Y. Li, C.~Zhang, and L.~Chen.
\newblock De-anonymizing social networks and inferring private attributes using
  knowledge graphs.
\newblock In {\em INFOCOM}, 2016.

\bibitem{radford2015unsupervised}
A.~Radford, L.~Metz, and S.~Chintala.
\newblock {Unsupervised representation learning with deep convolutional
  generative adversarial networks}.
\newblock {\em arXiv 1511.06434}, 2015.

\bibitem{rahmanmembership}
M.~A. Rahman, T.~Rahman, R.~Laganiere, N.~Mohammed, and Y.~Wang.
\newblock {Membership Inference Attack against Differentially Private Deep
  Learning Model}.
\newblock {\em Transactions on Data Privacy}, 2018.

\bibitem{salimans2016}
T.~Salimans, I.~Goodfellow, W.~Zaremba, V.~Cheung, A.~Radford, X.~Chen, and
  X.~Chen.
\newblock {Improved Techniques for Training GANs}.
\newblock In {\em NIPS}, 2016.

\bibitem{salimans2016weight}
T.~Salimans and D.~P. Kingma.
\newblock {Weight normalization: A simple reparameterization to accelerate
  training of deep neural networks}.
\newblock In {\em NIPS}, 2016.

\bibitem{shokri2015privacy}
R.~Shokri and V.~Shmatikov.
\newblock {Privacy-preserving deep learning}.
\newblock In {\em CCS}, 2015.

\bibitem{shokri2016membership}
R.~Shokri, M.~Stronati, C.~Song, and V.~Shmatikov.
\newblock {Membership inference attacks against machine learning models}.
\newblock In {\em IEEE Security and Privacy}, 2017.

\bibitem{song2017machine}
C.~Song, T.~Ristenpart, and V.~Shmatikov.
\newblock Machine learning models that remember too much.
\newblock In {\em ACM CCS}, 2017.

\bibitem{srivastava2014dropout}
N.~Srivastava, G.~E. Hinton, A.~Krizhevsky, I.~Sutskever, and R.~Salakhutdinov.
\newblock {Dropout: a simple way to prevent neural networks from overfitting.}
\newblock {\em Journal of machine learning research}, 2014.

\bibitem{theis2017lossy}
L.~Theis, W.~Shi, A.~Cunningham, and F.~Husz{\'a}r.
\newblock {Lossy image compression with compressive autoencoders}.
\newblock In {\em ICLR}, 2017.

\bibitem{tramer2016stealing}
F.~Tram{\`e}r, F.~Zhang, A.~Juels, M.~K. Reiter, and T.~Ristenpart.
\newblock {Stealing machine learning models via prediction apis}.
\newblock In {\em USENIX Security}, 2016.

\bibitem{triastcyn2018generating}
A.~Triastcyn and B.~Faltings.
\newblock Generating differentially private datasets using gans.
\newblock {\em arXiv preprint arXiv:1803.03148}, 2018.

\bibitem{demyst2018}
S.~Truex, L.~Liu, M.~E. Gursoy, L.~Yu, and W.~Wei.
\newblock {Towards Demystifying Membership Inference Attacks}.
\newblock {\em arXiv:1807.09173}, 2018.

\bibitem{verge}
J.~Vincent.
\newblock
  \url{https://www.theverge.com/2016/7/5/12095830/google-deepmind-nhs-eye-disease-detection},
  2016.

\bibitem{wainwright2012privacy}
M.~J. Wainwright, M.~I. Jordan, and J.~C. Duchi.
\newblock {Privacy aware learning}.
\newblock In {\em Advances in Neural Information Processing Systems}, 2012.

\bibitem{wu2015revisiting}
X.~Wu, M.~Fredrikson, W.~Wu, S.~Jha, and J.~F. Naughton.
\newblock {Revisiting differentially private regression: Lessons from learning
  theory and their consequences}.
\newblock {\em arXiv 1512.06388}, 2015.

\bibitem{wu2016automated}
X.~Wu and X.~Zhang.
\newblock {Automated Inference on Criminality using Face Images}.
\newblock {\em arXiv 1611.04135}, 2016.

\bibitem{wu2016quantitative}
Y.~Wu, Y.~Burda, R.~Salakhutdinov, and R.~Grosse.
\newblock {On the Quantitative Analysis of Decoder-Based Generative Models}.
\newblock In {\em ICLR (Poster)}, 2017.

\bibitem{yeh2016semantic}
R.~Yeh, C.~Chen, T.~Y. Lim, M.~Hasegawa-Johnson, and M.~N. Do.
\newblock {Semantic Image Inpainting with Perceptual and Contextual Losses}.
\newblock {\em arXiv 1607.07539}, 2016.

\bibitem{yeom2017unintended}
S.~Yeom, I.~Giacomelli, M.~Fredrikson, and S.~Jha.
\newblock {Privacy risk in machine learning: Analyzing the connection to
  overfitting}.
\newblock In {\em IEEE CSF}, 2018.

\end{thebibliography}

\appendix

\section{Unsuccessful Attacks}\label{app:unsuccessful}

\begin{figure*}[t]
   \centering
   \begin{minipage}[t]{.48\textwidth}
   \includegraphics[width=1\textwidth]{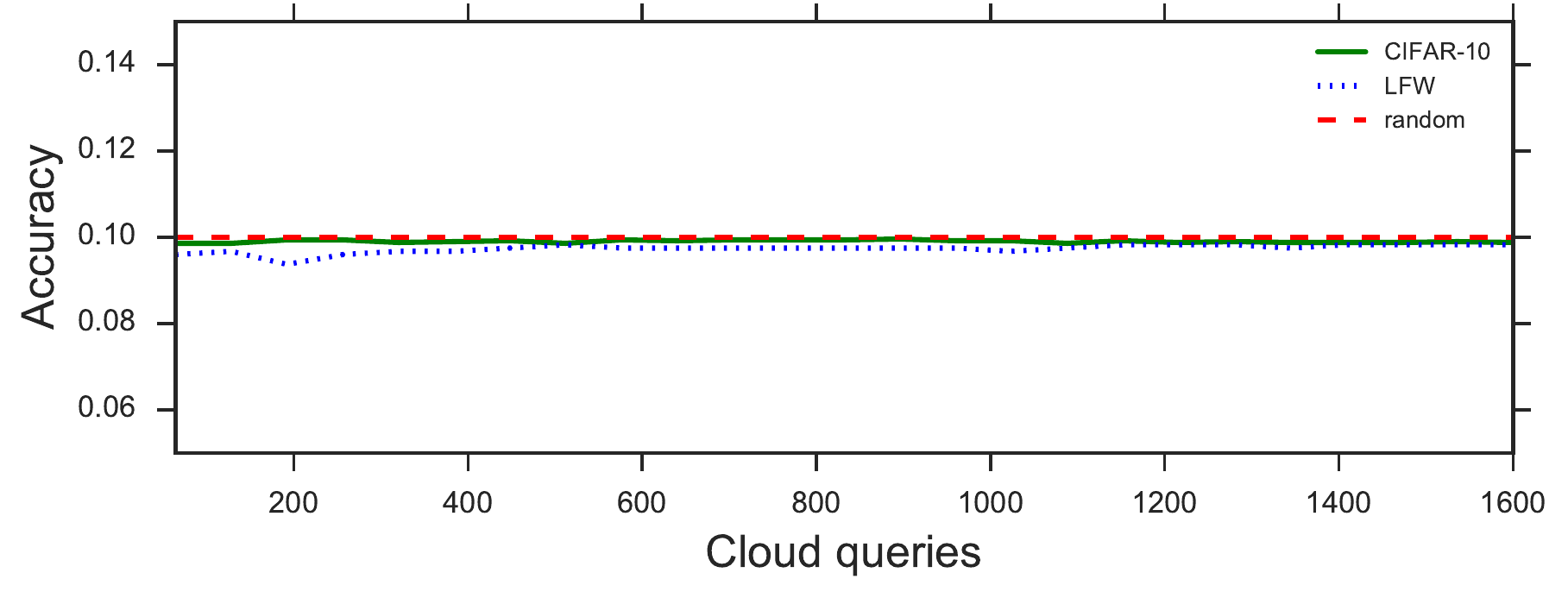}
   \vspace{-0.2cm}
   \caption{{Euclidean attack results for DCGAN target model trained on a random 10\% subset of CIFAR-10 and LFW.}}
   \label{fig:euclidean-attack-fig}
   \end{minipage}
~~~~
   \begin{minipage}[t]{.475\textwidth}
   \includegraphics[width=1\textwidth]{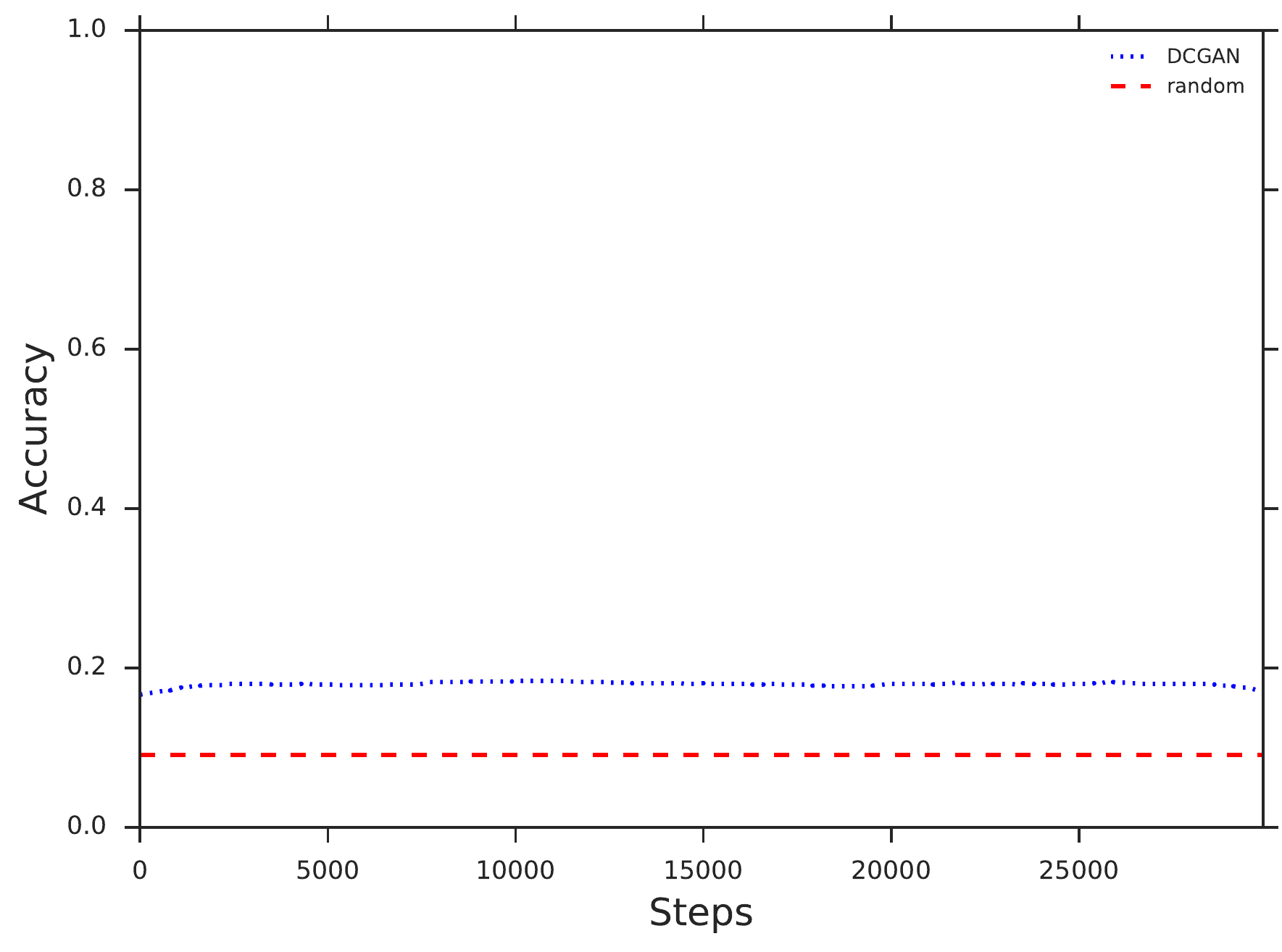}
   \vspace{-0.2cm}   
   \caption{{Black-box attack results with 10\% auxiliary attacker training set knowledge used to train a DCGAN \emph{shadow model} for DCGAN target model trained on a random 10\% subset of LFW.}}
   \label{fig:lfw-shadow}
   \end{minipage}
   \vspace{-0.2cm}   
\end{figure*}

We now report a few additional results, not included in the main body of the paper to ease presentation.
In Fig.~\ref{fig:euclidean-attack-fig}, we report the results of the Euclidean attack presented in~\ref{ssec:euclidean}.
This attack was performed on a target model (DCGAN) trained on a random 10\% subset of CIFAR-10 and 
a random 10\% subset of LFW, but we found that the attack did not perform much better than a random guess.

We also report on the results of a black-box setting where 10\% of training set samples from LFW are used to train a \emph{shadow model} -- see~Fig.~\ref{fig:lfw-shadow}.
Samples generated by this model are then injected into the attacker model together with the samples generated by the target model.
More specifically, at training time, each mini-batch is composed of synthetic samples generated either by the target model or by the shadow model.
However, this attack, inspired by the approach proposed by Shokri et al.~\cite{shokri2016membership}, only yields around 18\% of accuracy, with no improvements during training.

\section{Additional Samples}\label{app:samples}

\begin{figure*}[t]
   \centering
   \begin{subfigure}[b]{0.23\textwidth}
        \centering
        \includegraphics[width=\textwidth]{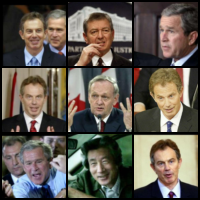}
        \caption{Real samples}
        \label{fig:real-lfw-top10-samples}
    \end{subfigure}
    ~~
    \begin{subfigure}[b]{0.23\textwidth}
        \centering
        \includegraphics[width=\textwidth]{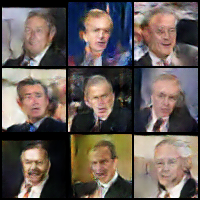}
        \caption{Target samples}
        \label{fig:cloud-dcgan-lfw-top10-samples}
    \end{subfigure}
    ~~
    \begin{subfigure}[b]{0.23\textwidth}
        \centering
        \includegraphics[width=\textwidth]{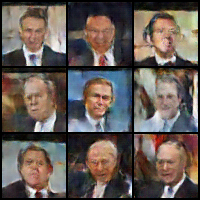}
        \caption{Attacker model samples}
        \label{fig:attacker-dcgan-cloud-dcgan-lfw-top10-samples}
    \end{subfigure}
    \caption{Various samples from the real dataset, target model, and black-box attack using the DCGAN target model on LFW, top ten classes.}
    \label{fig:wb-bb-samples-lfw-top10}
    \vspace{-0.2cm}
\end{figure*}

\begin{figure*}[t]
   \centering
   \begin{subfigure}[t]{0.24\textwidth}
        \centering
        \includegraphics[width=\textwidth]{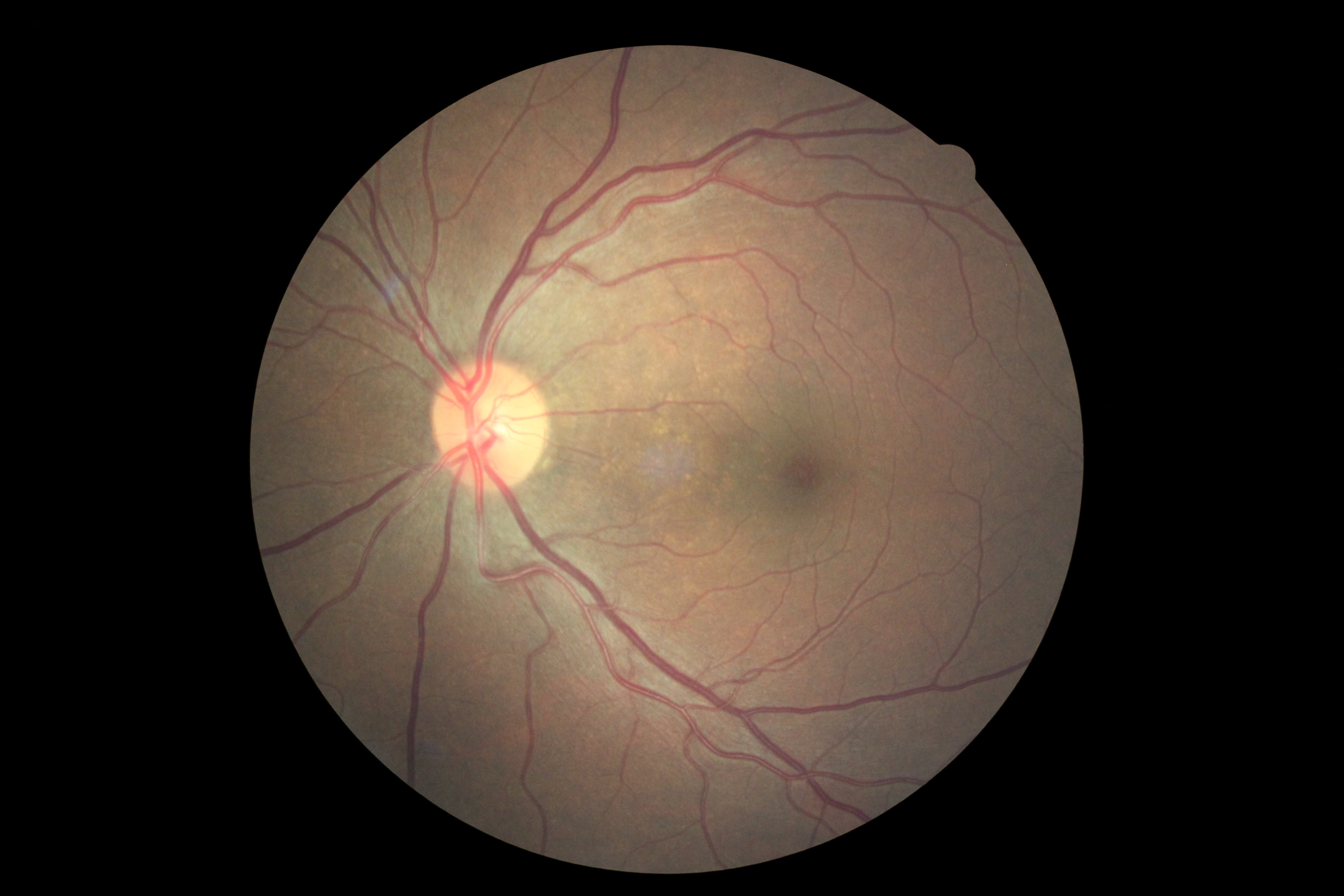}
        \caption{Real sample with no presence of diabetic retinopathy}
        \label{fig:no_diabetic_sample}
    \end{subfigure}
    ~~
    \begin{subfigure}[t]{0.24\textwidth}
        \centering
        \includegraphics[width=\textwidth]{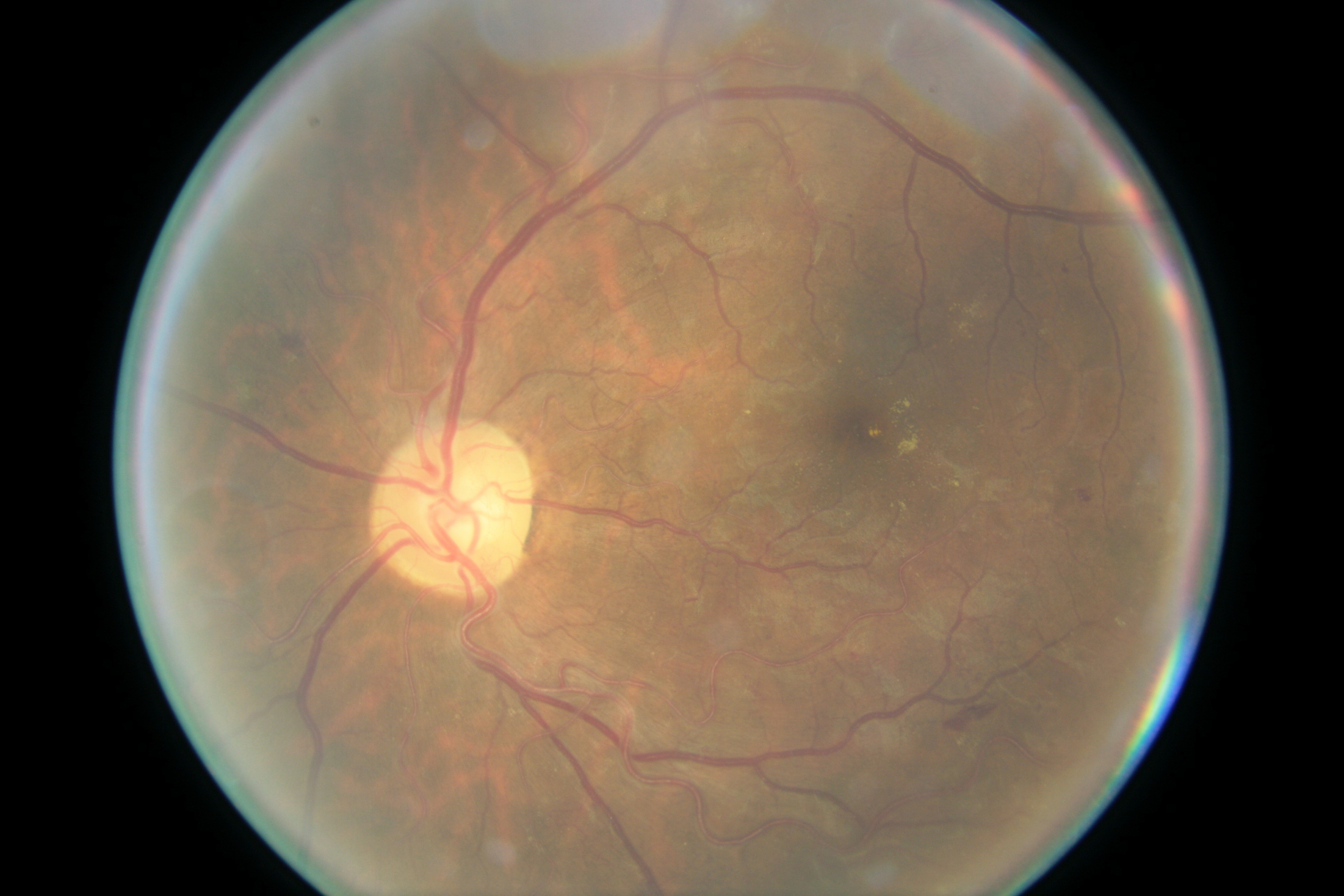}
        \caption{Real sample with high presence of diabetic retinopathy}
        \label{fig:diabetic_sample}
    \end{subfigure}
    ~~
    \begin{subfigure}[t]{0.47\textwidth}
        \centering
        \includegraphics[width=0.425\textwidth]{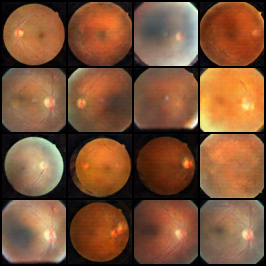}
        \caption{Selection of target generated samples classified with high confidence as belonging to the training set by both white-box and black-box attacks}
        \label{fig:cloud_diabetic_sample}
    \end{subfigure}
    \caption{Real and generated diabetic retinopathy dataset samples.}
    \label{fig:diab_ret_samples}
   \vspace{-0.2cm}    
\end{figure*}

\begin{figure*}[t]
   \centering
   \begin{subfigure}[b]{0.28\textwidth}
        \centering
        \includegraphics[width=\textwidth]{figures/lfw_top10_normalized_real_samples.png}
        \caption{LFW, top ten classes}
        \label{fig:real-lfw-top10}
    \end{subfigure}
    ~~
    \begin{subfigure}[b]{0.28\textwidth}
        \centering
        \includegraphics[width=\textwidth]{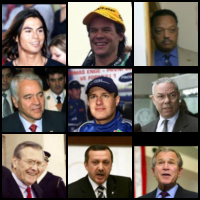}
        \caption{LFW, random 10\% subset}
        \label{fig:real-lfw-rand10}
    \end{subfigure}
    ~~
    \begin{subfigure}[b]{0.28\textwidth}
        \centering
        \includegraphics[width=\textwidth]{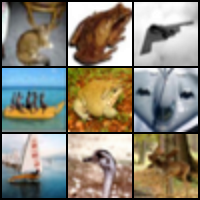}
        \caption{CIFAR-10, random 10\% subset}
        \label{fig:real-cifar-rand10}
    \end{subfigure}
   \vspace{-0.2cm}    
    \caption{Real samples.}
    \label{fig:real-samples}
   \vspace{-0.2cm}    
\end{figure*}

\begin{figure*}[t]
   \centering
   \begin{subfigure}[b]{0.28\textwidth}
        \centering
        \includegraphics[width=\textwidth]{figures/lfw_cloud_dcgan_top10_normalized_fake_samples_epoch_490.png}
        \caption{LFW, top ten classes}
        \label{fig:dcgan-lfw-top10}
    \end{subfigure}
    ~~
    \begin{subfigure}[b]{0.28\textwidth}
        \centering
        \includegraphics[width=\textwidth]{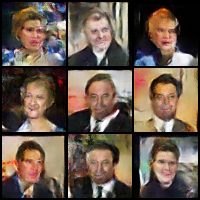}
        \caption{LFW, random 10\% subset}
        \label{fig:dcgan-lfw-rand10}
    \end{subfigure}
    ~~
    \begin{subfigure}[b]{0.28\textwidth}
        \centering
        \includegraphics[width=\textwidth]{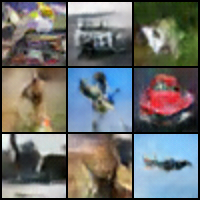}
        \caption{CIFAR-10, random 10\% subset}
        \label{fig:dcgan-cifar-rand10}
    \end{subfigure}
   \vspace{-0.2cm}    
    \caption{Samples generated by DCGAN target model.}
    \label{fig:dcgan-cloud-samples}
   \vspace{-0.2cm}    
\end{figure*}

\begin{figure*}[t]
   \centering
   \begin{subfigure}[b]{0.28\textwidth}
        \centering
        \includegraphics[width=\textwidth]{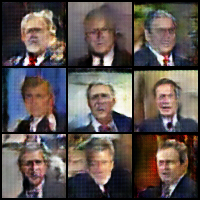}
        \caption{LFW, top ten classes}
        \label{fig:dcganvae-lfw-top10}
    \end{subfigure}
    ~~
    \begin{subfigure}[b]{0.28\textwidth}
        \centering
        \includegraphics[width=\textwidth]{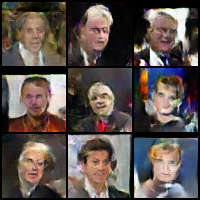}
        \caption{LFW, random 10\% subset}
        \label{fig:dcganvae-lfw-rand10}
    \end{subfigure}
    ~~
    \begin{subfigure}[b]{0.28\textwidth}
        \centering
        \includegraphics[width=\textwidth]{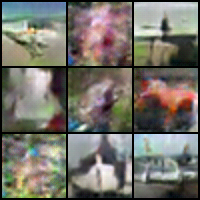}
        \caption{CIFAR-10, random 10\% subset}
        \label{fig:dcganvae-cifar-rand10}
    \end{subfigure}
   \vspace{-0.2cm}    
    \caption{Samples generated by DCGAN+VAE target model.}
    \label{fig:dcganvae-cloud-samples}
   \vspace{-0.2cm}    
\end{figure*}

\begin{figure*}[t]
   \centering
   \includegraphics[width=0.7\textwidth]{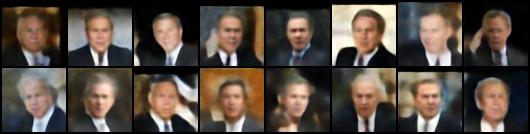}
   \vspace{-0.2cm}
   \caption{Samples generated by BEGAN target model on LFW, top ten classes.}
   \label{fig:began-lfw-top10}
   \vspace{-0.2cm}   
\end{figure*}

\begin{figure*}[t]
   \centering
   \includegraphics[width=0.7\textwidth]{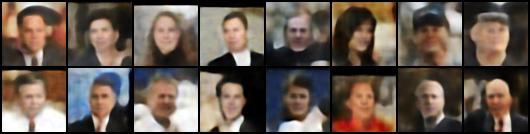}
   \vspace{-0.2cm}
      \caption{Samples generated by BEGAN target model on LFW, random 10\% subset.}
   \label{fig:began-lfw-rand10}
   \vspace{-0.2cm}
\end{figure*}

\begin{figure*}[t]
   \centering
   \begin{subfigure}[b]{\subb\textwidth}
       \centering
       \includegraphics[width=\textwidth]{figures/lfw_adv_dcgan_cloud_dcgan_top10_normalized_fake_samples_epoch_980.png}
       \caption{LFW, top ten classes}
       \label{fig:attacker-lfw-top10}
   \end{subfigure}
   ~~~
   \begin{subfigure}[b]{\subb\textwidth}
       \centering
       \includegraphics[width=\textwidth]{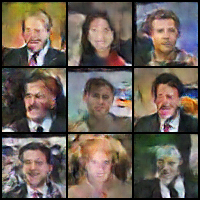}
       \caption{LFW, random 10\% subset}
       \label{fig:attacker-lfw-rand10}
   \end{subfigure}
   \vspace{-0.2cm}
   \caption{Samples generated by attacker model trained on samples from DCGAN target model on (a) LFW, top ten classes and (b) LFW, random 10\% subset.}
   \label{fig:attacker-samples}
   \vspace{-0.2cm}
\end{figure*}

In Figures~\ref{fig:wb-bb-samples-lfw-top10}--\ref{fig:attacker-samples}, 
we report additional examples of samples deferred from Section~\ref{sec:experiments}.
Specifically, 
real and generated samples are shown in Fig.~\ref{fig:wb-bb-samples-lfw-top10} for LFW and in Fig.~\ref{fig:diab_ret_samples} for the diabetic retinopathy (DR) dataset.
Then, Fig.~\ref{fig:real-samples} shows real samples from LFW and CIFAR-10, while Figures~\ref{fig:dcgan-cloud-samples}--\ref{fig:began-lfw-rand10} depict samples generated by various target models on LFW.
Finally, samples generated by the attacker model on LFW are reported in Fig.~\ref{fig:attacker-samples}.

\end{document}